\documentclass[aps,reprint,floatfix]{revtex4-2}
\usepackage{amsmath,amsfonts,amssymb}

\usepackage{graphicx}
\usepackage{bm}
\usepackage{mleftright}

\usepackage{mathtools}
\newcommand{\overbar}[1]{\mkern 2mu\overline{\mkern-2mu#1\mkern-2mu}\mkern 2mu}

\usepackage{float}

\usepackage{xcolor}
\definecolor{linkColor}{RGB}{0,70,120}
\definecolor{darkgreen}{RGB}{0,128,0}
\definecolor{darkgray}{RGB}{90,90,90}

\usepackage[colorlinks=true, allcolors=linkColor,pdfborder={0 0 0},pdfencoding = auto]{hyperref}

\usepackage[normalem]{ulem}

\begin{document}

\title{A Mean-Field Model for Active Plastic Flow of Epithelial Tissue}
\author{Nikolas H.\ Claussen}
\affiliation{Department of Physics, University of California Santa Barbara, Santa Barbara, California 93106, USA}
\author{Fridtjof Brauns}
\affiliation{Kavli Institute for Theoretical Physics, University of California Santa Barbara,
Santa Barbara, California 93106, USA}

\begin{abstract}
Animal morphogenesis often involves significant shape changes of epithelial tissue sheets. Great progress has been made in understanding the underlying cellular driving forces and their coordination through biomechanical feedback loops. However, quantitative understanding of how cell-level dynamics translate into large-scale morphogenetic flows remains limited.
A key challenge is finding the relevant macroscopic variables (order parameters) that retain the essential information about cell-scale structure.
To address this challenge, we combine symmetry arguments with a stochastic mean-field model that accounts for the relevant microscopic dynamics.
Complementary to previous work on the passive fluid- and solid-like properties of tissue, we focus on the role of actively generated internal stresses.
Centrally, we use the timescale separation between elastic relaxation and morphogenetic dynamics to describe tissue shape change in quasi-static balance of forces within the tissue sheet. The resulting geometric structure -- a triangulation in tension space dual to the polygonal cell tiling -- proves ideal for developing a mean-field model.
All parameters of the coarse-grained model are calculated from the underlying microscopic dynamics.
Centrally, the model explains how \emph{active plastic flow} driven by autonomous active cell rearrangements becomes self-limiting as previously observed in experiments and simulations.
Additionally, the model quantitatively predicts tissue behavior when coupled with external fields, such as planar cell polarity and external forces.
We show how such fields can sustain oriented active cell rearrangements and thus overcome the self-limited character of purely autonomous active plastic flow. These findings demonstrate how local self-organization and top-down genetic instruction together determine internally-driven tissue dynamics.
\end{abstract}

\maketitle

\section{Introduction}

Morphogenesis during animal development often relies on dramatic tissue reshaping through internal cell rearrangements. Prominent examples of this include convergent--extension shear flows during gastrulation \cite{Irvine.Wieschaus1994,Voiculescu.etal2007}, neurulation \cite{Elul.etal1997,Wallingford.Harland2002,Nikolopoulou.etal2017} and tubule formation \cite{Lienkamp.etal2012,Saxena.etal2014}. 
The dominant forces driving these tissue flows are generated between cells through controlled contractility along adherens junctions \cite{Bertet.etal2004,Fernandez-Gonzalez.etal2009} or through protrusions that pull on neighboring cells \cite{Shindo2018,Lecuit.etal2011}.
Phenomenological continuum models describe these driving forces as an active stress that is balanced against viscous dissipation \cite{Ishihara.etal2017,Streichan.etal2018,Saadaoui.etal2020,Serra.etal2021}.
However, how these tissue-scale dynamics and properties emerge from the underlying cellular behaviors remains a critical open question.

Most theoretical advances connecting the cell scale to the tissue scale \cite{Ishihara.etal2017,Czajkowski.etal2018,Hernandez.Marchetti2021,Duclut.etal2022,Grossman.Joanny2022,Fielding.etal2023} have focused on a particular class of vertex models with a passive area--perimeter elasticity \cite{Farhadifar.etal2007,Hufnagel.etal2007} which exhibits a fluid-to-solid transition \cite{Bi.etal2015}.
This postulated microscopic (i.e.\ cell scale) constitutive relation determines the passive rheological properties of the tissue onto which active driving forces (either cellular self-propulsion or active junctional contractility) are added.
Effective continuum theories based on the probability distribution of cell shapes capture certain features of the passive rheology of area--perimeter elasticity \cite{Ishihara.etal2017,Czajkowski.etal2018,Hernandez.Marchetti2021,Duclut.etal2022,Grossman.Joanny2022,Fielding.etal2023}.
However, the predictive power of these models for internally driven tissue flow has remained limited because the effect of active stresses is not derived from the cell scale but fitted \emph{a posteriori} to the macroscopic dynamics from cell-resolved simulations.

Motivated by recent experimental and theoretical findings \cite{Noll.etal2017,Noll.etal2020,Gustafson.etal2022,Brauns.etal2024}, we consider dominant active junctional tensions, while passive elastic/viscous forces are subdominant.
At short timescales (seconds), the mechanics of cells and their adherens junctions resemble those of a spring, where stress and strain are governed by a constitutive relationship.
However, over longer periods, the turnover of cytoskeletal components (actin, myosin, cadherins) leads to a decoupling of stress and strain from such relationships.
Instead, junctional tension is regulated by biomechanical feedback loops \cite{Guirao.Bellaiche2017}.
Negative feedback is required to stabilize the balance of forces within the tissue sheet \cite{Noll.etal2017,Gustafson.etal2022}.
Quasi-static remodeling of this force balance state, e.g.\ through positive feedback loops, can then drive tissue shape change through \emph{active plastic flow} \cite{Claussen.etal2023}.
An important feature of such self-organized internal remodeling is that the same cell-cell interfaces that drive cellular rearrangements are also annihilated and created through rearrangements. The result is a subtle interplay between the configuration of internal tensions and tissue flow that is not accounted for by existing continuum models.
This calls for the development of a new continuum description tailored for \emph{internally} driven epithelial tissue flow.

The fundamental challenge when moving to the continuum level is to identify the macroscopic degrees of freedom that retain the relevant information about the microscopic (cell-scale) structure. Identifying these order parameters is not possible based on symmetry arguments alone since many order parameters with the correct symmetries can be constructed from cell-scale features. Which of them are the relevant ones is not clear \emph{a priori}. Moreover, symmetry arguments do not constrain the coupling coefficients in the theory leaving them to be fitted to cell-resolved simulations \emph{a posteriori}.
To overcome these challenges, we combine symmetry arguments with bottom-up stochastic modeling at the cell scale that reveals the pertinent order parameters and their dynamic relationships.

\begin{figure}
    \centering
    \includegraphics{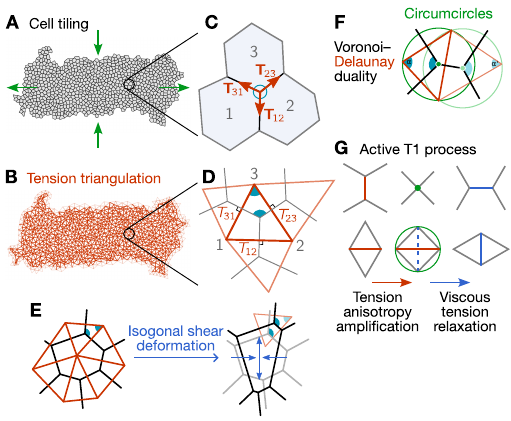}
    \caption{
    \textbf{A}~An epithelial sheet modeled as a polygonal tiling. Convergent extension appears as a large-scale shear flow (green arrows) driven by local cell rearrangements. 
    \textbf{B}~The tension triangulation dual to the cell tiling in force balance. Each node corresponds to a cell, and the length of the edge between two nodes encodes the active tension on the cell-cell interface.
    Remodeling of the tension triangulation drives tissue flow.
    \textbf{C}~At a tri-cellular vertex, the tensile forces $\mathbf{T}_{ij}$ have to sum to zero in force balance. 
    \textbf{D}~Rotating the tension force vectors constructs the tension triangle at a vertex. The vertex- and triangle angles are complementary.
    \textbf{E}~Isogonal deformations change cell shapes while keeping vertex angles (constrained by tension force balance) fixed. Isogonal deformations can create shear, and accommodate external or non-tension cell-internal forces.
    \textbf{F}~Voronoi--Delaunay construction of cell tiling by placing vertices at triangle circumcircles ensures vertex- and triangle angles are complementary. The angles $\alpha,\alpha'$ opposite to a cell-cell interface determine when it shrinks to length 0.
    \textbf{G}~Neighbor exchange (T1) process, driven by increasing tension on the central interface of a kite (pair of adjacent triangles). A T1 process corresponds to a flip in the tension triangulation. In the TDV model, it occurs when the triangle circumcircles coincide. Post-T1, tension relaxation leads to the elongation of a new cell-cell interface.
    }
    \label{fig:1}
\end{figure}

We model flat epithelial sheets as two-dimensional polygonal tilings of the plane (Fig.~\ref{fig:1}A). The central assumptions of our work are that the tissue is in quasi-static force balance and that the dominant stresses reside on the adherens-junctional cortices linking cells together~\cite{Lecuit.Yap2015,Agarwal.Zaidel-Bar2019}.
In this limit, the cortical tension forces sum to zero at each vertex. Geometrically, they form a triangle (see Fig.~\ref{fig:1}C). Since adjacent vertices share an edge, the tension triangles fit together to form a tissue-wide tension triangulation that is dual to the polygonal cell array \cite{Noll.etal2017} (Figs.~\ref{fig:1}A and~\ref{fig:1}B).
Morphogenetic dynamics through the adiabatic remodeling of tensions (by recruitment or detachment of motor molecules) can be formulated directly in terms of the tension triangulation \cite{Brauns.etal2024,Claussen.etal2023}.
Therein, the interplay between tension dynamics and cell rearrangements (T1 processes) is captured by simple geometric relationships, which we exploit to formulate a mean-field model. 

\subsection{Outline and summary of key results}

We begin by introducing a minimal cell-resolved model for tension-driven epithelial morphogenesis -- the \emph{tension-driven Voronoi} (TDV) model in Sec.~\ref{sec:TDV} -- which is based the tension triangulation and the dual Voronoi polygonal tiling.
In Sec.~\ref{sec:LTC}, we recapitulate the \emph{local tension configuration} parameters, which parameterize the shape of individual tension triangles and will form the basis for constructing macroscopic order parameters. 
For the specific winner-takes-all local feedback dynamics adopted here (Sec.~\ref{sec:local-tension-dyn}), we find that correlations between neighboring triangles due to the geometric compatibility constraints of the triangulation can be neglected. Moreover, correlations due to T1 transitions (edge flips in the triangulation) can be approximated using a mean-field approach based on the distribution of \emph{individual} tension triangle shapes and their orientation in space (Sec.~\ref{sec:mean_field}).
The leading moments of these statistical distributions provide the relevant fields (macroscopic order parameters) at the continuum level in a principled way (Sec.~\ref{sec:coarse_graining}). Their dynamics are derived from a mean-field ansatz together with a maximum-entropy approximation (Sec.~\ref{sec:T1-rate}).

The resulting mean-field model for active plastic flow reproduces key observations from experiments (\textit{Drosophila} germ-band extension \cite{Brauns.etal2024}) and cell-based simulations \cite{Claussen.etal2023}.
In these previous works, a mechanical feedback mechanism was proposed that leads to self-organized cell rearrangement and convergent--extension flow, oriented by the initial nematic order in the tension configuration.
It was shown that the self-organized T1s degrade nematic order, so that the flow is self-limiting, and the initial strength of order determines the total strain. The mean-field theory presented here analytically predicts this behavior and quantitatively matches cell-resolved simulations without parameter fitting.

In Sec.~\ref{sec:ext-fields}, we further extend this framework to account for coupling to external fields such as nematic patterns (motivated by the role of instructive genetic patterning during morphogenesis) as well as external tissue deformation and quantitatively predict the resulting tissue flow.
Both scenarios lead to sustained tissue flow, overcoming the self-limited character of purely autonomous active rearrangements.
In the case of externally imposed tissue deformation, the mean-field theory predicts a subtle shape-tension coupling that leads to a sustained strain rate \emph{perpendicular} to the axis of cell elongation. We validate this prediction with cell-resolved simulations. 
These results demonstrate the utility of the presented framework, which operates directly in the space of cortical tensions and provides new insight into the tissue-level control of morphogenetic flow.

\section{Tension-driven Voronoi model of quasi-static tissue shape change}{\label{sec:TDV}}

The starting point of our theory is the vertex-based tissue mechanics model introduced in Refs.~\cite{Brauns.etal2024,Claussen.etal2023} based on the assumption of tension-dominated force balance.
The central element of this model is the tension triangulation that represents the configuration of balanced tensions in the network of adherens junctions (Fig.~\ref{fig:1}B).
Specifically, force balance enforces that angles in the triangulation are complementary to those in the cell array as illustrated in Fig.~\ref{fig:1}D. However, these constraints leave soft isogonal (angle-preserving) modes that change cell shape while leaving the internal state of stress (i.e.\ the junctional tensions) invariant (Fig.~\ref{fig:1}E).
(Note that isogonal modes are distinct from the zero-modes of the ``area--perimeter'' vertex model \cite{Farhadifar.etal2007, Hufnagel.etal2007, Bi.etal2015} which arise because of excess target perimeter causing tensions to vanish \cite{Yan.Bi2019}).
The soft isogonal modes make it necessary to account for sub-leading contributions to the elastic energy due, for instance, to the resistance of the cell interior to shape distortions.
The total differential of elastic energy reads
\begin{align}\label{eq:total-energy}
     dE = \sum_{ij} T_{ij} d\ell_{ij} - p \sum_i dA_i + \varepsilon \sum_i dE_{\mathrm{shape},i}
\end{align}
where $i,j$ label cells, $T_{ij}$ and $\ell_{ij}$ are the interfacial tensions and lengths, $p$ is a constant pressure preventing collapse of the overall area $\sum_i A_i$, and the small parameter $\varepsilon$ multiplies the subdominant contributions from cell shape elasticity (see Appendix~\ref{app:simulation_methods} for the specific form of $E_{\mathrm{shape},i}$). The shape of the cell tiling in force balance is found by minimizing the elastic energy with respect to the vertex positions while holding the tensions $T_{ij}$ fixed.

Here, we propose a geometric \emph{ansatz} for the vertex positions in terms of the tension triangulation that automatically minimizes the dominant elastic energy: Given a tension triangulation, the Voronoi tessellation (constructed from the triangle circumcenters see Fig.~\ref{fig:1}F) yields a cell tiling where each interface is orthogonal to the corresponding triangulation edge. Therefore, such cell tilings automatically conform to the force balance of the junctional tensions, i.e.\ they minimize the leading ($\varepsilon^0$) terms in the elastic energy. For isotropic cell shapes, the Voronoi tessellation hence provides a convenient approximation of the elastic energy ground state and is compatible with experimental observations (see Appendix~\ref{app:simulation_methods}). We will utilize this simplified ``tension-driven Voronoi'' (TDV) model to develop a mean-field theory for tissue dynamics. It may also be of independent interest due to its computational simplicity.
In Sec.~\ref{sec:ext-fields}, show how this framework can be generalized to account for deviations from the Voronoi ``reference'' configuration due to isogonal deformations.

In the TDV model, tissue shape change results from the remodeling of the tension triangulation, i.e., \ changes of the tensions $T_{ij}$, biologically corresponding to redistribution and regulation of motor molecules on cell interfaces.
Large-scale tissue shear requires cell rearrangements (T1 transitions). These correspond to edge flips in the tension triangulation (Fig.~\ref{fig:1}G).
Delaunay--Voronoi duality implies that an edge flip happens when opposite angles $\alpha$ and $\alpha'$ in a kite (a pair of triangles sharing an edge) sum to $\pi$ (see Figs.~\ref{fig:1}F and~\ref{fig:1}G center):
\begin{equation}\label{eq:Delaunay}
    \alpha+\alpha' = \pi
\end{equation}
When this equality holds, the Voronoi vertices (the circumcircles of both triangles in the kite) coincide such that the cell-cell interface length vanishes.
This geometric \emph{Delaunay criterion} for cell rearrangements implies that the tension triangulation edge that flips is typically longer than its neighbors (i.e.\ opposite the largest angle in the triangle).
In other words, the cell-cell interface that collapses is typically the one under the highest tension.
This geometric insight motivates the concrete tension dynamics introduced in Sec.~\ref{sec:local-tension-dyn}.

\section{Order parameter for tension triangle shape}
\label{sec:LTC}

To derive an effective theory of this cell-scale model, we ``disassemble'' the tension triangulation into individual triangles, characterized by their probability distribution $\mathcal{P}$ within a mesoscopic tissue patch. The mesoscopic scale is taken much larger than the cell size but small enough to have no appreciable spatial gradients.
Due to tissue-wide force balance, the different tension triangles have to fit together to form a planar triangulation. This constraint implies that triangles are locally correlated.
However, we will argue that these correlations can be neglected, resulting in a simple mean-field model for the single-triangle distribution $\mathcal{P}$. From the leading moments of $\mathcal{P}$, we will construct large-scale order parameters, obtain their effective dynamics, and predict the evolution of tissue shape.

We begin by introducing a set of dynamical variables to describe the tension triangle shapes in a way suitable for coarse-graining. Clearly, this is not possible directly with the tensions and corresponding edge vectors. 
Instead, Refs.~\cite{Brauns.etal2024,Claussen.etal2023} introduced the \emph{local tension configuration} (LTC) parameters:
For a given tension triangle
we denote the three edges $i=1,2,3$ and the edge vectors  by $\mathbf{T}_i = (T_i^x, T_i^y)^\mathrm{T}$. Ordering the triangle sides by increasing length, we define the hexanematic matrix $\mathfrak{T}$ from the triangle edge vectors $\mathbf{T}_i$:
\begin{equation} \label{eq:Tfrak-def}
    \mathfrak{T} =
    \frac{1}{\sqrt{2}} \begin{pmatrix}
        T_1^x - T_2^x & T_1^y - T_2^y \\
        \sqrt{3} \, T_3^x & \sqrt{3} \, T_3^y 
    \end{pmatrix}.
\end{equation}
One can express $\mathfrak{T}$ in terms of its singular value decomposition (SVD)
\begin{equation}
    \mathfrak{T} = \mathbf{R}(\psi/3)\cdot \mathbf{M} \cdot \mathrm{diag}(\lambda_1, \lambda_2) \cdot \mathbf{R}(\phi)
\end{equation}
where $\mathbf{R}$ is a 2d rotation matrix of a given angle, and $\mathbf{M}=\mathrm{diag}\,(\pm 1, \pm 1)$ is a reflection matrix. $\mathfrak{T}$ can be thought of as the map from a reference equilateral triangle to the given tension triangle  (Figs.~\ref{fig:2}A and~\ref{fig:2}B). By the symmetries of the reference triangle ($2\pi/3$-rotation and reflection), we can always chose $\psi\in[0,\pi], \; \phi \in [0,\pi]$. 
The LTC parameters are the anisotropy magnitude $q$,  the ``shape phase'' $\psi$ which quantifies whether the triangle is obtuse or acute, and the triangle's moment of inertia axis in real space $\phi$. The anisotropy is defined from the singular values as 
\begin{equation}
    q = \frac{\lambda_2^2 - \lambda_1^2}{\lambda_2^2 + \lambda_1^2} \in [0, 1].
    \label{eq:LTC_definition}
\end{equation}
In passing, we note that $\psi$ is a ``hexanematic'' order parameter that describes the relative alignment of the nematic angle $\phi$ and the hexatic orientation of a tension triangle \footnote{The relevant orientation of the triangle has hexatic rather than triatic symmetry because a parity inversion leaves the nematic order parameter invariant.}.

As defined, $q$ and $\psi$ are invariant under spatial rotation, while $\phi$ has nematic symmetry. Under infinitesimal shear with magnitude $\delta s\ll q$ and principal axis orientation $\theta$, $q$ and $\phi$ transform as
\begin{equation}
    \delta \phi =  \sin\bigl(2(\theta-\phi)\bigr) \frac{\delta s}{2q}, \; \delta q = \cos\bigl(2(\theta-\phi)\bigr)\delta s
    \label{eq:LTC_transformation}
\end{equation}
We will later use this transformation law to describe the coupling of tension dynamics to external sources of anisotropy.
 
\section{Local tension dynamics}
\label{sec:local-tension-dyn}

The Delaunay criterion Eq.~\eqref{eq:Delaunay} implies that active cell rearrangements occur when the longest edge in a tension triangle grows compared to its neighbors (see Fig.~\ref{fig:1}G). 
After the neighbor exchange, the tension on the new edge has to decrease relative to its neighbors to allow T1 resolution, i.e.\ elongation of the new cell-cell interface (Fig.~\ref{fig:1}G). Without this relaxation, the system would locally get stuck in the four-fold vertex configuration~\cite{Bardet.etal2013}. 

\begin{figure*}
    \centering
    \includegraphics{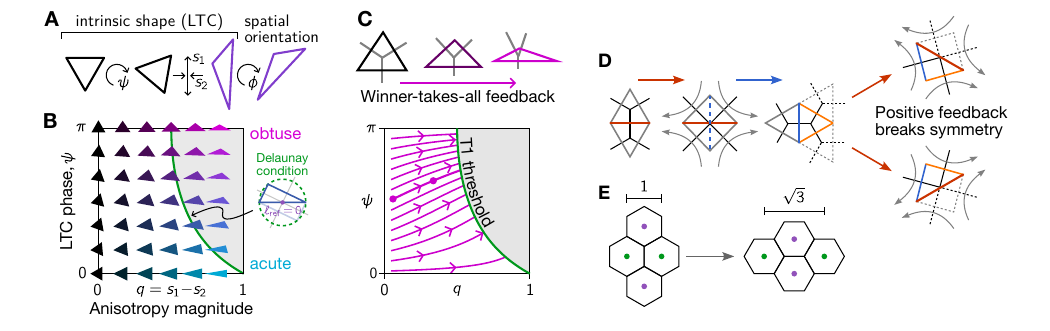}
    \caption{%
    \textbf{A}~Singular value decomposition of a triangle's shape.
    \textbf{B}~Intrinsic triangle shape is quantified by two parameters anisotropy $q$ and phase $\psi$ that span the LTC space. Triangles past the green T1 threshold violate the Delaunay condition for symmetric kites and are annihilated by triangulation edge flips (T1 events). 
    \textbf{C}~Positive feedback tension dynamics define a flow in LTC space, driving triangles towards the T1-threshold.
    \textbf{D}~The orientation of tension anisotropy after a T1 sensitively depends on the relative tensions before the T1. The edges highlighted in orange grow through tension feedback. The one that grows in tension fastest will cause the next T1. This post-T1 ``symmetry breaking'' degrades orientational order of tension anisotropy.
    \textbf{E}~Strain per T1, estimated in a symmetric lattice.}
    \label{fig:2}
\end{figure*}

The above considerations suggest the two basic ingredients required for locally self-organized cell rearrangement: positive feedback that increases the highest tension in each triangle to drive T1s, together with tension relaxation on newly formed edges.
Specifically, we consider tension dynamics driven by local positive feedback as proposed in \cite{Brauns.etal2024,Claussen.etal2023}, which was shown to reproduce many features of early \emph{Drosophila} convergent extension.
In the simplest form, such feedback can be formulated on the level of individual tension triangles.
For a tension triangle $\triangle$ with edge lengths $T_i, i\in\triangle$, we write the dynamics in a form that conserves the triangle perimeter $\sum_{i \in\triangle} \partial_t T_i = 0$
\begin{equation} \label{eq:tension-dyn}
    \tau \partial_t T_i(t) = f(T_i) - \frac{1}{3} \sum_{j \in \triangle} f(T_j),
\end{equation}
where $\tau$ is the overall timescale of tension dynamics \footnote{In Ref.~\cite{Claussen.etal2023}, we scaled $\tau = 25\:\mathrm{min}$ to match data from \emph{Drosophila} germ band extension}. This conservation is motivated by the limited pool of myosin in each cell.
We adopt a simple winner-takes-all feedback $f(T_i) = \overbar{T} (T_i/\overbar{T})^{n}$ parameterized by an exponent $n > 1$  (Fig.~\ref{fig:2}C, top). $\overbar{T}$ is the mean tension, whose absolute value is irrelevant in force balance.

The resulting dynamics can be visualized as a flow in triangle shape space (Fig.~\ref{fig:2}C, bottom). 
Triangle shapes flow to increasing anisotropy $q$, while the shape phase $\psi$ is only weakly affected:
\begin{equation} \label{eq:q-dyn}
   \tau \partial_t q(t) \approx c_q q, \quad \tau \partial_t \psi(t) \approx 0,
\end{equation}
where $c_q$ is a constant depending on the exponent $n$. Because of rotational symmetry, the orientation $\phi$ remains unchanged. Note that the exponent $n$ mainly affects the overall timescale $\tilde{\tau} = \tau/c_q$. For $n=4$, the value used in Ref.~\cite{Claussen.etal2023}, $c_q\approx 2.2$.
Equation~\eqref{eq:q-dyn} implies that positive feedback on tensions leads to the exponential growth of anisotropy. This growth is ``cut off'' by T1s, which happen along a critical line in LTC space (see Sec.~\ref{sec:mean_field} below).

After a cell neighbor exchange, the forces exerted on the new interface by its neighbors are generally not in mechanical equilibrium with the internal (myosin-generated) tension on the new interface.
A simple prescription for the handover of myosin during the cell neighbor exchange predicts that the external pulling forces always exceed the internal tension on a new interface (see Appendix~\ref{app:passive_tension}) so that it will extend until force balance is reestablished (Fig.~\ref{fig:1}G).
The timescale of this extension depends on the rate of viscous relaxation set by the turnover of passive crosslinkers. In addition, interface extension might induce the recruitment of myosin via strain rate feedback \cite{Noll.etal2017,Gustafson.etal2022}. 

So far, the described dynamics are triangle-autonomous and neglect that the triangles have to fit together to form a flat triangulation. 
In the TDV simulations, this constraint is implemented by ``flattening'' the triangulation in each timestep after the individual triangles have been updated (see Appendix~\ref{app:simulation_methods}).
This procedure forces the dynamics to evolve in the space of flat triangulations.
Correlations that arise from this flatness constraint are neglected in the mean-field model akin to the molecular chaos approximation in kinetic theory. This is justified because local positive feedback leads to rapid spatiotemporal decorrelation.
Another source of correlations of neighboring triangles are T1 transitions as they correspond to edge flips in the triangulation. 
How we deal with the resulting correlations is described in the next section.

\section{T1 transitions in mean-field approximation}
\label{sec:mean_field}

Under the Delaunay criterion Eq.~\eqref{eq:Delaunay}, an edge flip occurs when the opposing angles $\alpha, \alpha'$ in a kite sum to $\pi$.
Here, we assume that both $\alpha$ and $\alpha'$ are the largest angle in their respective triangle. In cell-resolved simulations, this is the case for over 90\% of edge flips since opposing angles in a kite are generally highly correlated (see Fig.~\ref{SI-fig:opposite_angles}).
Hence, \emph{on average}, triangles undergo a T1 when their maximal angle equals $\pi/2$. This defines the average T1 threshold $\mathcal{C}_\mathrm{T1}$ in the triangle shape space spanned by the LTC parameters $q$ and $\psi$ (see green line in Fig.~\ref{fig:2}B). To good approximation, the T1 threshold can be parameterized as $q_\mathrm{T1}(\psi)=\tfrac{1}{2}(1+(1-\psi/\pi)^3)$. Analysis of cell-resolved simulations shows that fluctuations around this average are generally weak since triangles in a kite share an edge and the opposing angles $\alpha,\alpha'$ are highly correlated (see Fig.~\ref{SI-fig:opposite_angles}). 
Figure~\ref{fig:2}C shows that the positive-feedback flow Eq.~\eqref{eq:tension-dyn} drives the local configuration of tensions towards the T1 threshold. Thus, active T1s in our model are driven by a local instability ``cut off'' by the T1 threshold.

In the mean-field picture in terms of the single-triangle probability distribution $\mathcal{P}$, T1 events correspond to the annihilation of two triangles and the creation of two new ones. 
The shape and orientation of the new triangles depend on the old ones pre-T1. In our mean-field model, we account for these ``scattering events'' by resetting a triangle $(q, \psi, \phi)$ when it hits the T1 threshold $q_\mathrm{T1}$.
Importantly, we subsume the viscous relaxation of tension on the new edge which reduces the anisotropy of the new triangles in the resetting:
$q$ is reset to a fixed constant $q_0$ that can be estimated from the minimum of $q$ after T1s in the full triangle dynamics (see Fig.~\ref{SI-fig:single_triangle_dynamics}). For the simulation parameters used in Refs.~\cite{Brauns.etal2024,Claussen.etal2023}, we find $q_0\approx 0.3$.
Moreover, during viscous stress relaxation, the shape phase $\psi$ fluctuates rapidly such that the shape phase of the new triangle is uncorrelated with the old one (see Fig.~\ref{SI-fig:single_triangle_dynamics}).
Lastly, the triangle orientation $\phi$ is stochastically deflected, setting the orientation of the triangle's next T1.
The random deflection results from decreasing tension on the newly formed interface and increasing tension on neighboring interfaces through positive feedback which causes a ``symmetry breaking'', as illustrated in Fig.~\ref{fig:2}D.
Starting from a nearly symmetric kite, this process amplifies small deviations from symmetry and therefore causes a random re-orientation.

In summary, the ``resetting rules'' are
\begin{subequations} \label{eq:resetting}
\begin{align}
    q &\mapsto q_0 \\ 
    \psi &\mapsto \mathrm{Uniform}(0,\pi) \\
    \phi &\mapsto \phi+\delta\phi, \; \delta\phi \sim \mathcal{N}(0,  \sigma_\phi^2)
\end{align}
\end{subequations}
where $\mathcal{N}(0,  \sigma_\phi^2)$ is the normal distribution with mean $0$ and variance $\sigma_\phi^2$. [Note that $\sigma_\phi^2$ is \emph{not} the variance of the orientation distribution $\mathcal{P}(\phi)$ but instead measures how much the local orientation of anisotropy is reoriented after a T1.] We can estimate $\sigma_\phi$ from the geometry of a T1: for a symmetric isosceles kite, the ``outer'' edges form a square at the point of the inner edge flip, implying that the new axis of tension anisotropy will be re-oriented by $\pm \pi/4$, providing the approximation $\sigma_\phi \approx \pi/4$. A more careful estimate (uniformly averaging over acute and obtuse triangles) yields $\sigma_\phi \approx \pi/5$ which we use in the following.

Together, the LTC flow Eq.~\eqref{eq:tension-dyn}, the T1 threshold $q_\mathrm{T1}(\psi)$, and the resetting rules Eq.~\eqref{eq:resetting} form a complete mean-field model for the dynamics of individual triangles $[q(t),\phi(t),\psi(t)]$.

T1s generate plastic strain, which can be estimated geometrically.
Figure~\ref{fig:2}E shows an idealized cell quartet with isotropic local tensions at the beginning and end of a T1 process. In the frame of reference aligned with the principal axes of tension anisotropy, the cell centroid positions change according to $\mathbf{x}_i \to \mathrm{diag} \, (\sqrt{3}, 1/\sqrt{3}) \cdot \mathbf{x}_i$. To evaluate the strain in the lab frame, we need to take into account the orientation of the kite. For any angle $\theta$ we define the nematic tensor as
\begin{align}
    \mathbf{N}(\theta) = \begin{pmatrix} \cos(2\theta) & \sin(2\theta) \\ \sin(2\theta) & -\cos(2\theta) \end{pmatrix}
\end{align}
The magnitude is written $N = |\mathbf{N}|$, where we use the convention $|N|^2 = \mathrm{Tr}\,(\mathbf{N}^\mathrm{T}\mathbf{N})/2$. A kite is composed of two triangles with orientations $\phi,\phi'$ not necessarily identical. To good approximation, the kite orientation is given by their circular average $\overbar{\{\phi, \phi'\}} = \arg\bigl(\tfrac12 e^{i2\phi}+ \tfrac12 e^{i2\phi'}\bigr)/2$, and hence the strain due to a T1 is
\begin{align}
    \mathbf{U}_\mathrm{T1}(\phi,\phi') = \exp\!\left[ \frac{\log 3}{2} \mathbf{N}\bigl( \overbar{\{\phi, \phi'\}} \bigr) \right]
    \label{eq:U_T1_kite}
\end{align}
As we will see below, fluctuations, i.e.\ $\phi\neq\phi'$, are important here. 

\section{Coarse-graining of mean-field dynamics}
\label{sec:coarse_graining}

The (stochastic) single-triangle mean-field model of the preceding section defines dynamics for the triangle distribution $\mathcal{P}$ which can be formulated as a Fokker--Planck equation (see Appendix~\ref{app:stochastic_model}). We now construct macroscopic order parameters and derive their dynamics by calculating the moments of $\mathcal{P}$. We use a bar over variables to denote averaged quantities. Details of the calculations below are presented in Appendix~\ref{app:coarse_graining}.

First, since the shape phase $\psi$ is randomized by T1s and since the flow Eq.~\eqref{eq:tension-dyn} has no strong $\psi$-bias, we assume that $\psi$ always remains uniformly distributed and thus does not appear on the hydrodynamic level.
[This approximation is specific for the tension dynamics Eq.~\eqref{eq:tension-dyn}. For other microscopic tension dynamics, $\psi$ may acquire non-trivial dynamics that would necessitate introducing a corresponding macroscopic order parameter.]
Further, the ``microscopic'' dynamics Eqs.~\eqref{eq:q-dyn} and~\eqref{eq:resetting} are invariant under spatial rotations and thus do not introduce any correlations between $q$ and $\phi$, which can therefore be considered independent. (In Sec.~\ref{sec:ext-fields}, we will show how coupling to an external field can give rise to correlations.)
The statistical independence of $\bar{q}$ and $\phi$ suggests that the averages $\bar{q}$ and $\overbar{\mathbf{N}}=\overbar{\mathbf{N(\phi)}}$ are the relevant macroscopic order parameters. By abuse of notation, we denote the magnitude of nematic order by $\overbar{N} = |\overbar{\mathbf{N}}|$, with the norm taken \emph{after} averaging. 
Notably, we \emph{do not} weight the nematic order by the magnitude of the anisotropy. $\overbar{\mathbf{N}}$ rather than $\overbar{\mathbf{Q}} =\overbar{q\mathbf{N}}$ determines strain due to T1s because positive feedback amplifies even weak anisotropy and the net anisotropy change during a T1-transition vanishes (Figs.~\ref{fig:2}C and~\ref{fig:2}E. Note that in Ref.~\cite{Claussen.etal2023} we quantified our simulations by measuring $\overbar{\mathbf{Q}}$). 

The rate of T1s, $\gamma$, is given by the probability flux integrated
across the T1-boundary $q_\mathrm{T1}(\psi)$, which can be expressed as~\footnote{
The factor $q$ appears because $q$ and $\psi$ are polar coordinates in the LTC space, so $q\, d q \, d\psi$ is the natural measure.}:
\begin{align}
    \gamma[\mathcal{P}] &= 
    \int_{\mathcal{C}_\mathrm{T1}}\!\!
    q d\mathbf{n} \cdot
    \begin{pmatrix} \dot{q} \\ \dot{\psi} \end{pmatrix} \, 
    \mathcal{P}(q,\psi,\phi)  \\
    &= \int\! q_\mathrm{T1}(\psi) d\psi \, \frac{\dot{q} - \dot{\psi} \, \partial_\psi q_\mathrm{T1}}{\sqrt{1 + \left(\partial_\psi q_\mathrm{T1}\right)^2}} \; \mathcal{P}[q_\mathrm{T1}(\psi), \psi, \phi],
    \label{eq:gamma_flux_integral}
\end{align}
where in the second line we have used an explicit parametrization for the T1 threshold curve $\mathcal{C}_\mathrm{T1}$ and its normal $\mathbf{n}$.
In the absence of external anisotropic fields, the T1 rate depends only on the intrinsic triangle shape distribution $\mathcal{P}(q, \psi)$ and not on spatial orientation $\phi$; an explicit expression will be provided below. 
For future reference, we define the $\psi$-averaged T1 threshold as $\overbar{q_\mathrm{T1}} = \pi^{-1}\int\! d\psi \, q_\mathrm{T1}(\psi) = 5/8$. 

By averaging Eqs.~\eqref{eq:q-dyn} and~\eqref{eq:resetting} one obtains (Appendix~\ref{app:coarse_graining}):
\begin{align}
    \partial_t  \bar{q}(t) &= \tilde{\tau}^{-1} \bar{q} - \gamma[\mathcal{P}] (\overbar{q_\mathrm{T1}} - q_0)
    \label{eq:q_dot} \\
    \partial_t \overbar{\mathbf{N}}(t) &= - \gamma[\mathcal{P}] 2 \sigma_\phi^2 \overbar{\mathbf{N}}. \label{eq:N_dot}
\end{align}
The mesoscopic strain rate $\overbar{\mathbf{U}}$ is given by averaging Eq.~\eqref{eq:U_T1_kite} where we assume that the two kite orientations $\phi, \phi'$ are drawn from $\mathcal{P}(\phi)$. Since the dynamics of $\phi$ is effectively a random walk (with a step taken at each T1 event), the distribution $\mathcal{P}(\phi)$ is a wrapped normal distribution, characterized by its standard deviation $\sigma = -\tfrac{1}{2} \log \bar{N}$, and we obtain
\begin{align}
    \partial_t \bar{\mathbf{U}}_\mathrm{T1} = \gamma[\mathcal{P}] \frac{\log 3}{2} h(\bar{N}) \bar{\mathbf{N}}
    \label{eq:U_dot}
\end{align}
where $h(\bar{N})$ is an approximation of the Gaussian integral over $\phi,\phi'$:
\begin{equation}
    h(\bar{N}) = \frac{|\overbar{\mathbf{N}\bigl[
    \overbar{\{\phi, \phi'\}}
    \bigr]}|}{
    \overbar{N}} \approx \frac{3}{2} -\frac{1}{2}\overbar{N}.
    \label{eq:kite_averaging}
\end{equation}
The strain rate is proportional to the T1-rate and oriented along the nematic order parameter $\overbar{\mathbf{N}}$. We verified Eqs.~\eqref{eq:N_dot} and \eqref{eq:U_dot} by substituting for $\gamma$ and $\overbar{\mathbf{N}}$ the rate and average orientation of T1 events tracked in full simulations, yielding good agreement (Fig.~\ref{SI-fig:assumptions_check}A).
In particular, this confirms that $\overbar{\mathbf{N}}$ is the relevant order parameter that determines the average strain by active T1s.

\begin{figure}[t]
    \centering
    \includegraphics{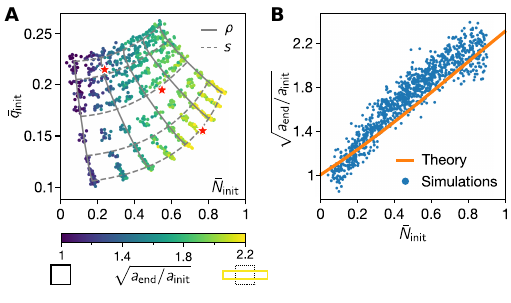}
    \caption{%
    Nematic order determines tissue shape change.
    \textbf{A}~Total aspect-ratio change in simulated tissue patches (reanalysis of simulations performed in Ref.~\cite{Claussen.etal2023}). Initial nematic order $N_\mathrm{init}$ and anisotropy magnitude $q_\mathrm{init}$ are controlled via the packing fraction $\rho$ and orientation bias $s$ of the protocol used to generate the initial condition (see main text). Solid and dashed gray lines show the ``coordinate grid'' induced by this parametrization. Red stars indicate the initial conditions for the time traces in Fig.~\ref{fig:4}.
    \textbf{B}~The mean-field theory predicts total strain with 90\% accuracy.
    }
    \label{fig:3}
\end{figure}

Equation~\eqref{eq:N_dot} predicts that $\overline{\mathbf{N}}$ will eventually decay to zero implying that large-scale tissue flow ceases after a finite amount of tissue shape change.
Since the T1 rate contributes to the strain rate in the same way as it does to the degradation rate of orientational order, we can combine Eqs.~\eqref{eq:N_dot} and~\eqref{eq:U_dot} to
\begin{align}
    d\overbar{\mathbf{U}}_\mathrm{T1} = -\frac{\log 3}{4 \sigma_\phi^2} h(\overbar{N}) \, d\overbar{\mathbf{N}}
\end{align}
The total strain (measured as the change of the aspect ratio, i.e.\ the ratio of eigenvalues of $\mathbf{U}$) is only a function of the initial order $\overbar{\mathbf{N}}(t=0)$, and independent of the T1 rate
\begin{equation} \label{eq:total-CE}
    \overbar{\mathbf{U}}_\mathrm{T1}(t\to\infty) = \exp\!\left[ \frac{\log 3}{4\sigma_\phi^2} H\bigl(\overbar{N}(0)\bigr) \overbar{\mathbf{N}}(0) \right]
\end{equation}
with $H(\overbar{N}) = \overbar{N}^{-1} \int_0^{\overbar{N}} \mathrm{d}\overbar{N}' \, h(\overbar{N}') =\tfrac{3}{2} - \tfrac{1}{4}\overbar{N}$. 

A key prediction from Eq.~\eqref{eq:total-CE} is that the total strain does not depend on the initial magnitude of anisotropy $q$, but only on the nematic order $\overbar{\mathbf{N}}$. The reason for this is that positive feedback amplifies the magnitude of tension anisotropy.
To check this prediction, we re-plot the data from the ``phase diagram'' of Ref.~\cite{Claussen.etal2023} as a function of the initial values of $\bar{q}$ and $\overbar{N}$ (see Fig.~\ref{fig:3}).
The total strain can be measured as the change of the aspect ratio $a$ (i.e.\ the ratio of eigenvalues of $\overbar{\mathbf{U}}_\mathrm{T1}$). Initial tension triangulations were generated from hard-disk packings with packing fraction $\rho$. The resulting isotropic triangulations were then stretched by a pure shear with magnitude $s$ to induce tension anisotropy. Varying the control parameters $\rho$ and $s$ maps out the order parameter plane for initial average anisotropy magnitude $\bar{q}(0)$ and magnitude of nematic order $\overbar{N}(0)$ as shown in Fig.~\ref{fig:3}A. 
Clearly, the resulting total extension of the tissue patch (indicated by color) depends very little on $\bar{q}_\mathrm{init}$ and quantitatively matches the prediction from the phenomenological theory, Eq.~\eqref{eq:total-CE}. Indeed, Eq.~\eqref{eq:total-CE} has a 90\% correlation with the values found in the full simulations (see Fig.~\ref{fig:3}B).

\begin{figure}
    \centering
    \includegraphics{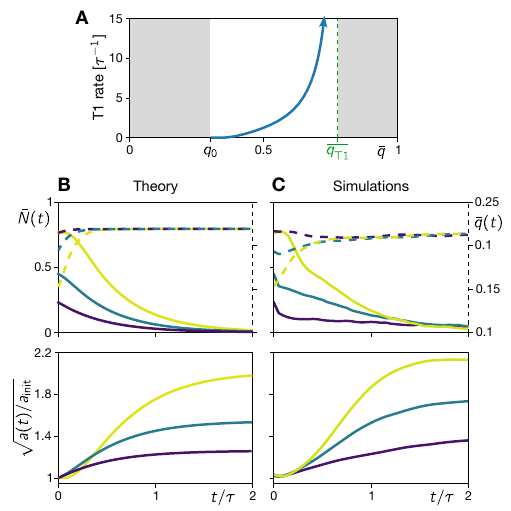}
    \caption{
    Mean-field theory captures dynamics of cell-resolved simulations.
    \textbf{A}~T1-rate as a function of mean anisotropy in maximum-entropy approximation. The T1-rate diverges at $\overbar{q}$ approaches the T1 threshold $\overbar{q_\mathrm{T1}}$.
    \textbf{B--C}~Comparison of dynamics in the mean-field model (B) and full cell-resolved simulations (C) performed in Ref.~\cite{Claussen.etal2023} for three different initial conditions corresponding to the red stars in Fig.~\ref{fig:3}C. (Each trajectory in C shows an average of 12 realizations.)
    }
    \label{fig:4}
\end{figure}

\section{T1 rate in maximum entropy approximation}
\label{sec:T1-rate}

Let us now turn to the transient dynamics, which depend on the T1 rate $\gamma$. Equation~\eqref{eq:gamma_flux_integral} defines the T1 rate in terms of the probability flux across the T1 threshold. This integral depends, in principle, on the details of the distribution $\mathcal{P}(q, \psi)$, i.e.\ the higher moments $\overbar{q^2}, \overbar{q^3}$, etc. 
The key qualitative behavior of $\gamma$ is, however, independent of these details. For $\bar{q} \ll \overbar{q_\mathrm{T1}}$, T1s are rare, i.e.\ $\gamma \approx 0$. As anisotropy grows, probability mass shifts towards the T1 threshold, increasing $\gamma$ until the positive and negative contributions in Eq.~\eqref{eq:q_dot} balance so that $\bar{q}$ approaches a fixed point $\bar{q}_* =  \tilde{\tau}\gamma_*(\overbar{q_\mathrm{T1}} - q_0)$.
Notably, $\bar{q}_*$ is independent of the positive feedback timescale $\tilde{\tau}$ since the $\tilde{\tau}$-dependence cancels between the faster anisotropy growth and the T1 rate, $\gamma \propto \tilde{\tau}^{-1}$ that drives degradation of $q$.

To obtain an approximation for $\gamma$ in terms of only the first moment of $\mathcal{P}$ we use the method of maximum entropy with the constraints that (1) the marginal of $\psi$ is uniform, (2) the mean of $q$ is $\bar{q}$, (3) $q$ always lies in the ``legal'' region of shape space, $q_0\leq q \leq q_\mathrm{T1}(\psi)$ (see Appendix~\ref{app:max_ent}). Substituting the maximum entropy distribution parameterized by the mean anisotropy $\bar{q}$ into the flux integral Eq.~\eqref{eq:gamma_flux_integral} yields the T1 rate $\gamma(\bar{q})$ plotted in Fig.~\ref{fig:4}A.
For $\bar{q} = q_0$, the maximum entropy distribution becomes $\delta(q-q_0)$ and $\gamma$ vanishes.
Conversely, $\gamma$ diverges for $\bar{q} \to \overbar{q_\mathrm{T1}}$.

It is important to keep in mind that the ``average T1 threshold'' $\overbar{q_\mathrm{T1}}$ and the approximation for $\gamma$ provided here are valid only for a uniform distribution of the LTC phase $\psi$ which results from winner-takes-all positive feedback. In contrast, saturating positive feedback generates tension cables ($\psi \approx 0$) \cite{Claussen.etal2023}, so that $\overbar{q_\mathrm{T1}} \approx 1$ and $\gamma$ would be strongly suppressed for $\bar{q} < 1$.

Using the maximum-entropy approximation for $\gamma(\bar{q})$, one can now integrate Eq.~\eqref{eq:q_dot} to obtain $\bar{q}(t)$. As expected, $\bar{q}$ simply converges to the fixed point by $\bar{q}_*$. The (numerical) solution for $\bar{q}(t)$ can be substituted into Eq.~\eqref{eq:N_dot} to find the time evolution of nematic order $\overbar{\mathbf{N}}(t)$. Good qualitative agreement with the full simulations is achieved as shown in Figs.~\ref{fig:4}B and~\ref{fig:4}C.
Importantly, a single set of effective parameters $(q_0, \sigma_\phi)$ reproduces the dynamics across the entire range of initial conditions $\bar{q}(0), \overbar{N}(0)$. In the following, we will show how this minimal model can be extended to account for the coupling to external fields.

\section{Coupling to external fields}
\label{sec:ext-fields}

So far, we considered purely autonomous tension dynamics where the rate of change of tensions only depends on the local tensions themselves. 
Yet morphogenesis typically involves the interplay of multiple factors: local mechanical feedback, instructive genetic patterning, and external forces acting at the boundaries of the tissue.
Such factors can be included in our framework as external fields as we shall show in the following. We consider two cases: (A) an anisotropic modulation of tension regulation and (B) tissue deformation driven by forces other than cortical tensions, e.g.\ external stresses or active bulk stresses in the cell interior.

\subsection{Sustained source field for tension anisotropy}

\begin{figure}
    \centering
    \includegraphics{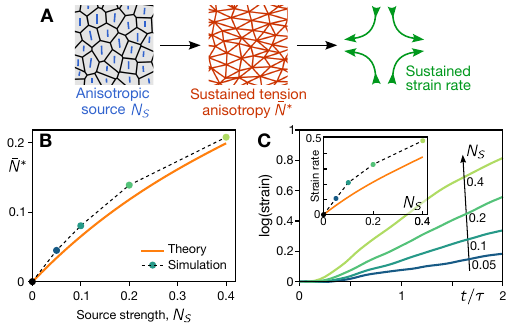}
    \caption{
    \textbf{A}~Illustration of an ``external'' anisotropic field, e.g.\ planar cell polarity, (blue sticks) that couples to the local tension dynamics and thus induces anisotropic tension, resulting in oriented T1s that drive shear flow. 
    \textbf{B}~Steady-state nematic order versus the strength of the anisotropic source shows good agreement between full simulations and the mean-field theoretical prediction.
    \textbf{C}~Strain as a function of time from numerical simulations for various anisotropic source strengths $N_\mathrm{S}$. 
    Inset: Steady-state strain rate (fitted in the interval $t = 0.5 - 1.5$ versus the strength of the anisotropic source.
    (Each data point corresponds to an average of three stochastically sampled initial conditions.)
    }
    \label{fig:tension-source}
\end{figure}

In many biological contexts, such as primitive streak formation \cite{Voiculescu.etal2007}, the \textit{Drosophila} notum \cite{Bosveld.etal2012}, and \textit{Xenopus} neurulation \cite{Butler.Wallingford2018} there is strong evidence that nematically ordered patterns of protein localization, such as planar cell polarity (PCP) \cite{Shindo.Wallingford2014}, influence cortical tension dynamics.
Moreover, external forces can cue the internal tension regulation via mechanical feedback responding to the strain rate \cite{Noll.etal2017}, for example preceding \textit{Drosophila} germ band extension \cite{Gustafson.etal2022}.
Both situations can be captured by coupling tension dynamics to an ``external'' nematic source, illustrated by the blue sticks in Fig.~\ref{fig:tension-source}A.
Such coupling to external nematic fields is often assumed in continuum models \cite{Ishihara.etal2013,Popovic.etal2017} and has recently been investigated in the context of a vertex model with area--perimeter elasticity \cite{Duclut.etal2022}.
In the following, we show that thanks to positive feedback in our model, even a weak external source is sufficient to orient self-organized T1s and maintain persistent coherent tissue flow.

We parameterize the  orientation and strength of the external source of anisotropy by a prescribed nematic tensor $\mathbf{N}_\mathrm{S}$ and modify the tension dynamics Eq.~\eqref{eq:tension-dyn} to account for the amplification of tension along edges aligned with the source
\begin{equation}
    \label{eq:tension-dyn-source}
    f(T_i) = \overbar{T}(T_i/\overbar{T})^{n} + (\mathbf{e}_i^\mathrm{T} \cdot \mathbf{N}_\mathrm{S} \cdot\mathbf{e}_i) T_i
\end{equation}
where $\mathbf{e}_i$ is the unit vector along edge $i$.
Even when starting from a disordered initial condition, the coupling to the source field will induce nematic order in the tensions $\overbar{\mathbf{N}} \propto \mathbf{N}_\mathrm{S}$ that is maintained in steady state by a balance of the anisotropic source and T1-driven degradation.
To derive the source term in the mean-field equation for the order parameter $\overbar{\mathbf{N}}$, first consider a single tension edge. In an infinitesimal time interval $dt$, the anisotropic source will cause an additional tension increment $dT_i \propto (\mathbf{e}_i^\mathrm{T} \cdot \mathbf{N}_\mathrm{S} \cdot\mathbf{e}_i) T_i \, dt$ due to the anisotropic source. Geometrically, this means that each tension triangle will undergo an infinitesimal shear transformation $(\mathbf{N}_\mathrm{S}/2) dt$, rotating its principal axis [see Eq.~\eqref{eq:LTC_transformation}]:
\begin{equation} \label{eq:phi_dot_biased}
    \tau \partial_t \phi =  \frac{N_\mathrm{S}}{2q} \sin(2(\phi_S - \phi)),
\end{equation}
where $\phi_S$ is the orientation of source anisotropy. The factor $1/q$ comes in because triangles close to equilateral (small $q$) are more easily reoriented by shear than those with a strong existing axis of anisotropy (large $q$).

Averaging over the single-triangle distribution $\mathcal{P}$ yields the modified mean-field dynamics (see Appendix~\ref{app:coarse_graining})
\begin{equation} \label{eq:n_dot_biased}
    \partial_t \overbar{\mathbf{N}} = - 2\sigma_\phi^2 \gamma \overbar{\mathbf{N}} + \frac{1}{2\bar{q} \tau} (1-\overbar{N}^4) \mathbf{N}_\mathrm{S},
\end{equation}
which, compared to autonomous tension dynamics, Eq.~\eqref{eq:N_dot}, has an additional term proportional to $\mathbf{N}_\mathrm{S}$.
The factor $(1-\overbar{N}^4)$ arises because shear-induced rotation vanishes for $\phi=\phi_S$ and thus ensures that nematic order saturates at a maximum magnitude of $1$. Mathematically, it results from the Gaussian integral over $\phi$ (see Appendix~\ref{app:coarse_graining}).

In the limit of weak source strength $N_\mathrm{S} \ll 1$, Eq.~\eqref{eq:n_dot_biased} predicts steady-state nematic order $\overbar{\mathbf{N}}_* =  \mathbf{N}_\mathrm{S}/(4\sigma_\phi^2 \bar{q}_* \tau \gamma(\bar{q}_*))$, in good agreement with numerical simulations across a large range of source strength $N_\mathrm{S}$ (see Fig.~\ref{fig:tension-source}B).
As a result of the sustained nematic order, active T1s acquire an orientational bias leading to a sustained strain rate according to Eq.~\eqref{eq:U_dot}, again in good agreement with numerical simulations as shown in Fig.~\ref{fig:tension-source}C.
Phenomenologically, this scenario corresponds to a viscous flow (Stokes) driven by an internal active stress.
However, the rate of strain here is not set by the balance of stress and viscous dissipation but by the rate of active internal remodeling, i.e.\ by feedback acting on tensions. 
In passing, we note that the magnitude of anisotropy in steady state, $\bar{q}_*$, is not affected by the anisotropic source. This is because the growth rate of $q$ and the T1 rate $\gamma$ are proportionally influenced by $N_\mathrm{S}$ and therefore cancel in the steady state [cf.\ Eq.~\eqref{eq:q_dot}].

Taken together, this analysis shows how the ``external'' source of tension anisotropy allows the tissue to overcome the limitation of purely autonomous local tension dynamics. We anticipate that this is important in biological systems such as primitive streak formation \cite{Voiculescu.etal2007,Serra.etal2023}, where flow is sustained despite disorder in the cell packing.

\subsection{Isogonal strain couples cell shape and tension anisotropy}

So far, we have exclusively considered the dynamics of cortical tensions and their consequence for tissue shape change while neglecting isogonal deformations.
While they leave the instantaneous tension state invariant, isogonal modes can influence the tension dynamics, notably by shifting when interfaces undergo T1 transitions.
In the following, we extend our mean-field model to describe the effect of small isogonal strain.

\begin{figure}
    \centering
    \includegraphics{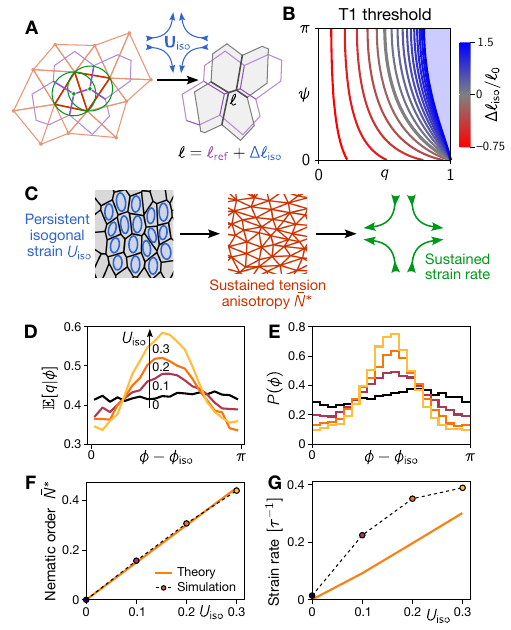}
    \caption{
    \textbf{A}~Isogonal (angle-preserving) deformations can collectively deform cells compared to the Voronoi reference state while maintaining junctional force balance.
    \textbf{B}~Isogonal contributions $\Delta\ell_\mathrm{iso}$ to interface lengths shift the T1 threshold, with more anisotropy required to collapse isogonally stretched interfaces. 
    \textbf{C}~The interplay of isogonal strain, which causes a direction-dependent shift of the T1 threshold, and local tension dynamics leads to a sustained tension anisotropy (D, E and F) and thus results in sustained flow (G).
    \textbf{D}~Orientation-conditional average anisotropy for different levels of persistent isogonal strain.
    \textbf{E}~Probability distribution of tension orientations for different levels of persistent isogonal strain.
    \textbf{F}~Steady-state nematic order $N_*$ versus magnitude of persistent isogonal strain: theoretical prediction (solid orange line) vs observation in simulations (disks). 
    \textbf{G}~Strain rate versus magnitude of persistent isogonal strain: theoretical prediction (solid orange line) vs observation in simulations (disks).
    Each data point corresponds to an average over three stochastically sampled initial conditions.
    }
    \label{fig:6}
\end{figure}

Following Ref.~\cite{Brauns.etal2024}, we define isogonal strain $\mathbf{U}_\mathrm{iso}$ through the displacement of the cell centroids relative to their Voronoi-reference positions given by the triangulation vertices, as illustrated in Fig.~\ref{fig:6}A. Since we consider an incompressible tissue we are primarily interested in a purely deviatoric isogonal strain.
We limit our analysis to small isogonal strain as a perturbation to active tension dynamics. In this limit, tension dynamics still determines which interface in a given kite will undergo a T1, but the isogonal strain modifies \emph{when} the T1 occurs because it changes interface lengths and thus shifts the T1 threshold (see Figs.~\ref{fig:6}A and~\ref{fig:6}B).
Interfaces aligned with $\mathbf{U}_\mathrm{iso}$ are stretched while those that are orthogonal shrink, relative to their Voronoi-reference length. The shift of the T1 threshold is therefore orientation dependent: A higher tension anisotropy is required to drive the collapse of an interface aligned with the direction of isogonal stretching.
To linear approximation, we can write this as a shift $\delta \overbar{q_\mathrm{T1}} \propto \mathrm{Tr}(\mathbf{N}\cdot\mathbf{U}_\mathrm{iso})$ (see Appendix~\ref{app:isogonal_fokker_planck}).
As a consequence of the orientation-dependent T1 threshold, the intrinsic properties of the tension triangles, parameterized by $(q, \psi)$ become correlated with the triangles' orientations in real space $\phi$. To measure this correlation in simulations (or data), we will use the conditional average of anisotropy $\mathbb{E}[q|\phi]$.

As a concrete example, we simulate cells with intrinsically elongated shapes \cite{Comelles.etal2021} which can result from boundary forces or from active stresses in the cell interior \cite{Lin.etal2022}.
This cell elongation results in a persistent isogonal strain $\mathbf{U}_\mathrm{iso}$, which causes $\mathbb{E}[q|\phi]$ to become biased (see Fig.~\ref{fig:6}D) as the orientation-dependent T1 threshold delays T1s oriented along $\mathbf{U}_\mathrm{iso}$, allowing more anisotropy to build up along this direction.
In addition, the T1s that take place at large anisotropy lead to reduced reorientation of anisotropy after the edge flip (see Appendix~\ref{app:coarse_graining}).
This results in an increase of $\mathcal{P}(\phi)$ along the direction where $\mathbb{E}[q|\phi]$ is largest.
Together these effects cause nematic order $\overbar{\mathbf{N}}$ to build up along $\mathbf{U}_\mathrm{iso}$ (see Fig.~\ref{fig:6}E).
In turn, T1s become oriented yielding a net strain rate \emph{perpendicular} to the axis of cell elongation (see Fig.~\ref{fig:6}G).

Remarkably, these effects are quantitatively predicted by the mean-field model of Sec.~\ref{sec:mean_field} without additional phenomenological parameters [see Appendix~\ref{app:isogonal_fokker_planck}, where these results are derived using the Fokker--Planck equation for $\mathcal{P}(\phi)$].
While the induced nematic order $\overbar{\mathbf{N}}^*$ is predicted quite precisely, Eq.~\eqref{eq:U_dot} underestimates the strain rate. This is likely because in Eq.~\eqref{eq:U_dot} for the T1-strain we assumed isotropic cell shapes (cf.\ Fig.~\ref{fig:2}E) neglecting the influence of isogonal deformations on cell shape.

So far, we have considered a patch of tissue with free boundaries. As a complementary example, consider a piece of tissue with \emph{pinned} boundaries.
In this case, the total strain rate must vanish, such that cell rearrangements are compensated by cell elongation through isogonal strain, $\partial_t \mathbf{U} = \partial_t \overbar{\mathbf{U}}_\mathrm{T1} + \partial_t \mathbf{U}_\mathrm{iso} = 0$.
In general, isogonal strain accommodates tissue deformations driven by forces at its boundary.
Crucially, while isogonal strain was held steady above, here it increases dynamically, driven by the tension dynamics. As a result, the T1 threshold becomes a ``moving target,'' continuously shifting towards higher tension anisotropy as T1s occur. Since $\partial_t \mathbf{U}_\mathrm{iso} \propto \overbar{\mathbf{N}}$, the T1 threshold increases most for triangles aligned with $\overbar{\mathbf{N}}$. This mechanism causes tissue flow to stall. 
In numerical simulations with fixed boundaries, we indeed observe a 40\% reduction of the T1 rate and increasing $\bar{q}$ and $\overbar{N}$ as isogonal strain builds up (see Fig.~\ref{SI-fig:fixed_boundaries}). \emph{Drosophila} \emph{Toll[RM9]} mutants during germ-band extension are a biological example of this scenario \cite{Irvine.Wieschaus1994,Claussen.etal2023}. The \emph{Toll[RM9]} mutation abolishes the soft region abutting the contracting germ band, pinning the germ band boundaries and suppressing large-scale tissue flow,  while highly anisotropic active tension (myosin) is maintained  \cite{Irvine.Wieschaus1994} as cells rearrange locally \cite{Lefebvre.etal2023}.
A quantitative description of this scenario in the mean-field theory will require dynamical treatment of large isogonal strain, going beyond the steady state analysis performed for small isogonal strain above, and is an interesting avenue for future work.

\section{Discussion}

We have derived a mean-field model for the dynamics of internally generated stress of an epithelial tissue sheet in approximate quasi-static force balance.
The adiabatic dynamics are not governed by a microscopic constitutive relationship but instead result from biomechanical feedback loops that act on the cell scale and actively remodel the internal stress state.
Specifically, we considered local positive feedback acting on junctional tensions, which drives convergent--extension as an \emph{active plastic flow}.

The basis of our approach lies in characterizing the mechanical state of the tissue via the marginal distribution of individual tension triangle shapes -- representing tension configurations at tri-cellular vertices. Because the winner-takes-all positive-feedback dynamics considered here rapidly degrade spatial correlations, the information contained in the single-triangle distribution is sufficient to quantitatively describe the active plastic flow of tissue.
We anticipate that this approximation will be insufficient to describe active \emph{ordering} of tissue, as observed e.g.\ in the \emph{Drosophila} pupal wing where cells rearrange into a regular hexagonal packing \cite{Classen.etal2005}, which exhibits strong spatial correlations.

Our theory reproduces the inherent cross-coupling between the nematic order of active tension and self-organized cell rearrangements observed in experiments \cite{Brauns.etal2024} and cell-resolved numerical simulations \cite{Claussen.etal2023}:
Tension anisotropy drives and orients T1s which, in turn, degrade the nematic order of tension as the highest tension interfaces are annihilated in T1 processes.
Thus, tissue flow by autonomous, self-organized T1s sensitively depends on initial order and is self-limiting. 
Sustained flow requires persistent reservoirs of nematic order, which we incorporate through external fields in our model. Examples of such sources of anisotropy are ``hoop stresses'' due to the shape of an embryo \cite{Lefebvre.etal2022} and planar cell polarity established by biochemical signaling pathways \cite{Voiculescu.etal2007,Bosveld.etal2012,Shindo.Wallingford2014}.
Moreover, cell shape change can feed back into the tension dynamics through strain rate recruitment \cite{Noll.etal2017,Gustafson.etal2022} and thereby act as an anisotropic ``external'' source (see Appendix~\ref{app:strain_rate_feedback}).
This effect is similar to flow (or strain-rate) alignment in liquid-crystal systems, where the microscopic constituents (such as rods or disks) align to the strain rate of the surrounding fluid. There, flow alignment depends on the shape of the constituents. In contrast, the alignment of nematic order $\overbar{\mathbf{N}}$ in tension space does not automatically couple to physical strain (rate) since isogonal deformations can change cell shape while leaving the tension configuration invariant. Instead, strain (rate) alignment is governed by biomechanical feedback loops.

We have shown that persistent cell elongation can also induce nematic order of tension and thus result in oriented T1s.
Shape-tension coupling is commonly assumed in phenomenological models \cite{Rozman.etal2023,Comelles.etal2021,Doostmohammadi.etal2018,Ishihara.etal2017}.
In our model, it emerges naturally from the interplay of the positive feedback that drives T1s and the viscous relaxation of stress that completes the T1 process after a cell neighbor exchange has occurred.
This suggests that shape-tension coupling may not require specific feedback loops but can arise generically from feedback-based regulation of internally generated stress. 
Remarkably, the resulting tissue strain rate is oriented \emph{perpendicular} to the cell elongation.
A similar phenomenon was found in a phenomenological continuum model implementing stress-based positive feedback \cite{Ioratim-Uba.etal2023} and in vertex model simulations where cell elongation was implemented through a particular cell-shape elastic energy \cite{Duclut.etal2022}. In this vertex model, T1 transitions are driven by stochastic fluctuations of junctional tensions, in contrast to the positive-feedback-driven dynamics studied here.
Since we have formulated a stochastic model as a bridge from the deterministic cell-scale dynamics, Eq.~\eqref{eq:tension-dyn}, to the continuum, we expect that including stochastic fluctuations in the cell-scale dynamics will be straightforward.

When bridging from the cell scale to the tissue scale, it is crucial to identify the relevant macroscopic order parameters. 
Previous coarse-graining approaches have been based on cell shape in physical space \cite{Marmottant.Graner2007,Hernandez.Marchetti2021,Fielding.etal2023}.
In contrast, our framework is based on the tissue's internal state of stress.
Junctional tensions, rather than their lengths, are the degrees of freedom directly controlled by cells via motor and adhesion molecules.
At the local level of tri-cellular vertices, balanced junctional forces are described in terms of tension-triangle shapes. 
We have formulated a stochastic model for these triangle shapes capturing the positive-feedback-driven local tension dynamics. A mean-field approximation of this stochastic model yielded the relevant macroscopic order parameters:
A scalar $\bar{q}$, describing the average magnitude of anisotropy, and a nematic tensor $\bar{\mathbf{N}}$ describing the average orientation of anisotropy. 
Interestingly, the separation of anisotropy magnitude and orientation into two order parameters has also been employed to describe cell shape in physical space \cite{Hernandez.Marchetti2021,Fielding.etal2023}.
While $\bar{q}$ and $\bar{\mathbf{N}}$ are sufficient to describe the dynamics for autonomous local positive feedback, additional order parameters become relevant when the dynamics are coupled to external fields or when the feedback dynamics are modified.
In the former case, we found that correlations between local anisotropy magnitude $q$ and orientation $\mathbf{N}$ become important.
These correlations can be encoded in an additional nematic field
\begin{align}
    \overbar{q \mathbf{N}}-\bar{q}\overbar{\mathbf{N}} = \int\!d\phi \, \mathcal{P}(\phi) \, \big(\mathbb{E}[q|\phi] - \bar{q}\big) \,\mathbf{N},
\end{align}
which would need to be incorporated in future extensions of our mean-field model to describe the effect of isogonal strain dynamically, not just in a uniform steady state.
 
The hexanematic alignment parameter $\psi$ that distinguishes between different local configurations of anisotropic tension \cite{Brauns.etal2024} has dropped out of our mean-field model because of the specific choice of winner-takes-all tension feedback.
Other feedback mechanisms may strongly bias $\psi$, making it a relevant dynamical variable on the mean-field level, represented, for instance, by the first Fourier component $\Psi$ of its probability distribution $\mathcal{P}(\psi) = 1/\pi + \Psi \cos{\psi}$. The case of \emph{saturating} positive feedback will be particularly interesting to further investigate since it drives $\psi$ to zero \cite{Claussen.etal2023}, implying the formation of tension cables that suppress cell rearrangements \cite{Monier.etal2010,Umetsu.etal2014,Sknepnek.etal2023}.
Going forward, it will be interesting to explore possible relations between the hexanematic order in tension space and in physical space \cite{Armengol-Collado.etal2023}.

An important limitation of the presented model is that it describes a uniform ``mesoscopic'' patch of cells without large-scale gradients. Incorporating such gradients will be necessary to develop a complete continuum model.
Gradients of the active strain will generically lead to geometric incompatibility, which, in a flat tissue, must be compensated by isogonal deformations that restore compatibility of the total strain while maintaining local force balance.
The relationship between geometric incompatibility (i.e.\ curvature) in tension space and isogonal modes will be the subject of a forthcoming manuscript. 
Here we have described the effect of small isogonal strain on active T1s.
This perturbative description will not work for large isogonal strain that causes additional passive T1s \cite{Brauns.etal2024} similar to cell rearrangements in a (two-dimensional) foam \cite{Weaire.etal2005,Marmottant.etal2008,Raufaste.etal2010}.
Passive T1s relax internal stresses and allow the tissue to flow plastically in response to external forces.
We anticipate that the yield threshold for passive T1s will depend on local nematic order as well as the orientation of local hexatic order, similar to the situation in foams \cite{Kraynik.etal1991}.
Geometric incompatibility in the plane can also be resolved by out-of-plane deformations \cite{Murisic.etal2015,Fuhrmann.etal2023}, suggesting that biological tissue can employ active T1s for ``shape programming'' \cite{Duffy.Biggins2020,Resetic2024}.

\acknowledgements{
We thank Boris Shraiman and Cristina Marchetti for insightful discussions and feedback on the manuscript. NHC was supported by NIGMS R35-GM138203 and NSF PHY:2210612. FB acknowledges support of the GBMF post-doctoral fellowship (GBMF award \#2919). 
This work was partly funded by a grant from ICAM, the Institute for Complex Adaptive Matter, to NHC.
}

\appendix

\section{Viscous stress relaxation}
\label{app:passive_tension}

In addition to the tension dynamics Eq.~\eqref{eq:tension-dyn}, the cell resolved model of Ref.~\cite{Claussen.etal2023} contains a mechanism that ensures tension relaxation after a T1: A new edge starts with a low level of active tension determined by a previously proposed ``handover'' of motor molecule concentration \cite{Brauns.etal2024}. The total tension on a new edge is the sum of the active tension $T^{(a)}$ and the viscous stress $T^{(v)}$ (called ``passive tension'' in Refs.~\cite{Brauns.etal2024,Claussen.etal2023}), which models viscous stresses and decays with rate $\tau_v^{-1}$: $\partial_t T^{(v)} = -\tau_v T^{(v)}$. Viscous stresses are required to balance the pulling forces on the new edge because the initial motor molecule level is low.
Interface extension might also induce an increase of active tension through strain-rate induced myosin recruitment \cite{Noll.etal2017,Gustafson.etal2022}.
In the minimal model considered here, we do not account for this effect.

\begin{figure}
    \centering
    \includegraphics[width=0.8\linewidth]{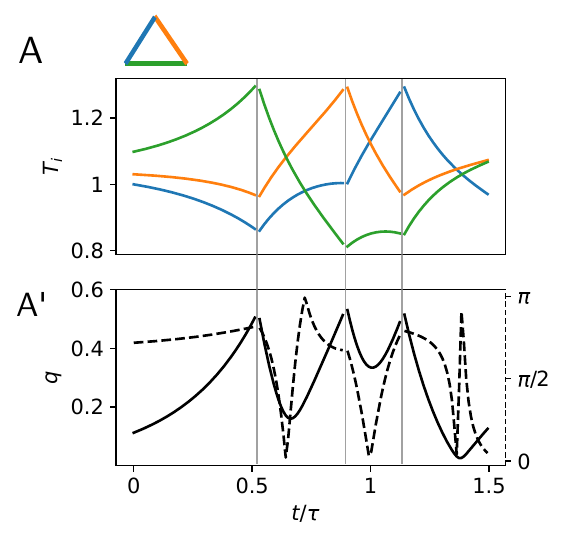}
    \caption{
    Single-triangle dynamics in space of tensions and LTC parameters.
    \textbf{A} Dynamics of a symmetric kite constructed from a single tension triangle due to positive feedback and viscous stress relaxation. Parameters as in Ref.~\cite{Claussen.etal2023}. 
    T1s occur at the symmetric Delaunay threshold $\alpha=\pi/2$, and the triangle post-T1 is constructed by assuming a point-symmetric kite, i.e.\ a kite made out of two identical tension triangles.
    In this case, the circumcircle matching condition ensures the edge lengths are continuous across T1s.
    \textbf{A'} Corresponding dynamics of the LTC parameters. Note the rapid fluctuation of $\psi$ post T1: as the triangle edge lengths intersect, the triangle changes from obtuse to acute. This is taken into account by our assumption that $\psi$ is randomized post T1.}
    \label{SI-fig:single_triangle_dynamics}
\end{figure}

Let us illustrate this dynamics for a symmetric tension kite, i.e. a kite made from two identical triangles $(T_1, T_2, T_3)$. In a symmetric kite, T1 transitions occur when the maximal angle in the triangle $=\pi/2$. Because of the circumcircle condition, the edge lengths of the new triangle post-flip agree with the edge lengths pre-flip: $(T_1, T_2, T_3) \mapsto (T_1, T_2, T_3)$. The handover mechanism predicts an initial level of active tension $T^{(a)}_{1} = T_2+T_3-T_1$, and hence a viscous stress $T^{(v)}_{1}=T_1 - T^{(a)}_{1}$ immediately post T1.
Fig.~\ref{SI-fig:single_triangle_dynamics}A shows the result of the positive feedback tension dynamics combined with viscous stress relaxation:  the initially highest tension increases due to positive feedback, eventually leading to a T1 when the Delaunay criterion $\alpha \leq\pi/2$ is violated. Post-T1 relaxation of the viscous stress drives down the tension on the flipped edge over a time $\sim\tau_v$, and the process begins anew. The ``reset values'' in the mean-field model of Sec.~\ref{sec:mean_field} correspond to those we expect a time $\sim \tau_v$ after the T1-event.
We can also see that the shape phase $\psi$ fluctuates rapidly after a T1: as the triangle edge lengths intersect, the triangle changes from obtuse to acute. This is taken into account in the mean-field model by the randomization of $\psi$.

\begin{figure}
    \centering
    \includegraphics[width=0.8\linewidth]{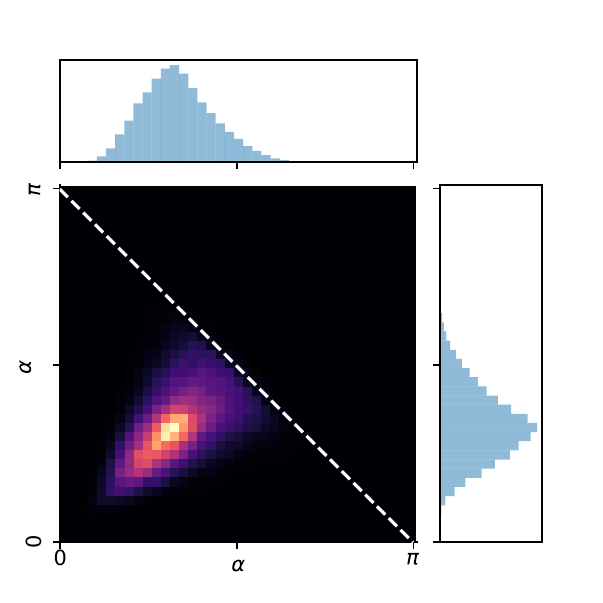}
    \caption{
    2D histogram of triangle angles $\alpha,\alpha'$ that are opposite of a shared edge (see Fig.~\ref{fig:1}C), from the TDV simulation shown in Fig.~\ref{SI-fig:assumptions_check}.
    By triangle geometry, the mean of $\alpha,\alpha'$ is $\pi/3$, and by the Delaunay criterion, $\alpha+\alpha' < \pi$.
    }
    \label{SI-fig:opposite_angles}
\end{figure}

\section{Additional information on mean-field model}{\label{app:stochastic_model}}

\subsection{Estimate of $\sigma_\phi$ from triangle geometry}{\label{app:D_phi_estimate}}

To estimate the degree of re-orientation of a triangle by a T1, we assume that the orientation post-T1 is given by that of the second-longest edge in the triangle pre-T1 -- as discussed in the main text, this is the edge that will ``win'' in the tension feedback and undergo a T1 next. For a perfectly obtuse triangle $\psi=\pi$, this yields a reorientation by $\pm \pi/4$. To average over $\psi$, we use the steady-state solution of the stochastic model for LTC dynamics presented below. Averaging of the second-longest edge of a triangle at the T1 threshold using the steady-state distribution of the mean-field model yields the estimate $\sigma_\phi=0.2\pi$.

Above, we assumed that T1s occur when the maximum triangle angle is $\alpha=\pi/2$. But the same calculation can be done for the case of a shifted T1 threshold, as occurs in the presence of isogonal strain. We return to this below in Appendix \ref{app:isogonal_fokker_planck}.

\subsection{Steady-state distribution of marginals}{\label{app:steady_state}}

In the steady state, triangles are reset to $\psi\sim\mathrm{Uniform}(0,\pi)$ post T1, and take a time
\begin{align}
    \tau_\mathrm{T1} = \tilde{\tau} \log(q_\mathrm{T1}/q_0)
    \label{eq:tau_T1}
\end{align}
until the next T1. Because $q_\mathrm{T1}(\psi))$ is a decreasing function of $\psi$, this leads to two subtle effects: the marginal distribution of all triangles is slightly biased towards $\psi=0$, $\mathcal{P}(\psi)\propto \tau_\mathrm{T1}$. 
This bias is indeed observed in the full simulations Fig.~\ref{SI-fig:Voronoi_vs_elastic} and even in the empirical data from the \emph{Drosophila} embryo\cite{Brauns.etal2024}.
The distribution of triangles undergoing a T1 is however biased towards $\psi=\pi$, $\mathcal{P}(\psi | q=q_\mathrm{T1}(\psi))\propto \tau_\mathrm{T1}^{-1}$.
In the calculations in the main text, we ignore these weak biases and treat $\psi$ as uniformly distributed.

Regarding the distribution of $q$, we have $\dot{q}=q$ and hence $\mathcal{P}(q|\psi) = \left(q \log(q_\mathrm{T1}(\psi)/q_0)\right)^{-1}$. Averaging over $\psi$, we have  $\mathcal{P}(q) \approx \left(q \log(\overbar{q_\mathrm{T1}}/q_0)\right)^{-1}$ where $\overbar{q_\mathrm{T1}}$ is the average T1 threshold. In particular, the mean is $\bar{q}_* = (\overbar{q_\mathrm{T1}}-q_0)/\log (\overbar{q_\mathrm{T1}}/q_0)$. This shows that the parameter $q_0$ only has a weak influence, as long as $0<q_0 < \overbar{q_\mathrm{T1}}$.

\subsection{Fokker--Planck equation for mean-field dynamics}

The mean-field model of Sec.~\ref{sec:mean_field} can be formulated as a Fokker--Planck equation for the time-dependent probability distribution $\mathcal{P}(q, \psi, \phi)$:
\begin{equation}
    \partial_t \mathcal{P} = - \nabla \cdot \mathbf{J} + s = -\nabla \cdot ( \begin{pmatrix} \dot{q} \\ \dot{\psi} \\ \dot{\phi} \end{pmatrix} \mathcal{P}) +  \nabla^2 (D \mathcal{P}) + s
\end{equation}
The flux $\mathbf{J}$ comprises a drift term due to the deterministic tension dynamics and a diffusion term due to noise.
In the following, we restrict ourselves to the case $D = 0$, where the dynamics in between T1s is purely deterministic.
T1s annihilate triangles and generate new ones. They thus appear as an absorbing boundary at $q=q_\mathrm{T1}(\psi)$, i.e.\ $\mathcal{P} |_{q=q_\mathrm{T1}(\psi)}=0$, and a source $s$ that re-injects triangles post-T1. The T1-rate is the integral of the probability flux $\mathbf{J}$ across the T1 boundary: $\gamma = \int_{\mathcal{C}_\mathrm{T1}} \! q d\mathbf{n} \cdot \mathbf{J} $.
Note that the factor $q$ in the integral appears because $q$ and $\psi$ are polar coordinates in the LTC space. This is because the intrinsic shape of a triangle is represented by the traceless part of the symmetric matrix $\mathfrak{T}^\mathrm{T} \cdot \mathfrak{T}$ [cf.\ Eq.~\eqref{eq:Tfrak-def}] that can be parameterized by the complex order parameter $qe^{i\psi}$.

By balance of probability, we must have $\int_{q,\psi,\phi} \gamma = \int_{q,\psi,\phi} s$. The form of the source is determined by our resetting procedure Eq.~\eqref{eq:resetting}:
\begin{equation}
    s(q, \psi, \phi) = \frac{\delta(q-q_0)}{q_0 \pi} \cdot \int_{q',\psi',\phi'} \frac{e^{-\frac{(\phi-\phi')^2}{2\sigma_\phi^2}}}{\sqrt{2\pi \sigma_\phi^2}} \gamma(q',\psi',\phi')
\end{equation}
The factor $(q_0 \pi)^{-1}$ accounts for the uniform spreading along the $\psi$-axis, and the convolution of the T1-flux with a Gaussian kernel in $\phi$ accounts for the reorientation.
This equation could be easily adapted to different post-T1 dynamics.

\begin{figure*}
    \centering
    \includegraphics[width=\textwidth]{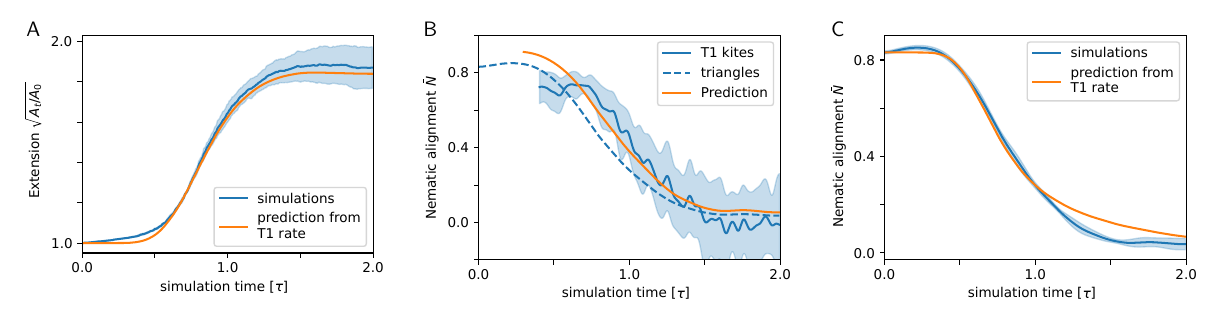}
    \caption{
    Key predictions of the mean-field model match full cell-resolved simulations.
    \textbf{A}~Oriented T1 rate predicts tissue-scale shape change. T1-rate and mean T1-orientation taken from the numerical simulation.
    To compute the orientation of T1-kites, we extracted all T1-events at a given time point from the simulation and computed the orientation of the flipped triangulation edge. Note that in the initial phase of the simulation, there are very few T1s, and hence the T1-orientation is not defined.
    \textbf{B}~Nematic alignment of tension triangles and T1-kites. Prediction based on Eq.~\eqref{eq:U_dot}.
    \textbf{C}~Mean orientation of tension triangle, and fit based on Eq.~\eqref{eq:N_dot}, with T1 rate taken from simulation. Note this fit uses the same $\sigma_\phi^2 = 0.2\pi$ as in the main text.
    All panels show averages over 3 simulations with stochastic initial conditions.}
    \label{SI-fig:assumptions_check}
\end{figure*}

\section{Additional information on coarse-graining calculation}{\label{app:coarse_graining}}

\subsection{Averaging over triangle orientation $\phi$}{\label{app:phi_averaging}}

\begin{figure}[t]
    \centering
    \includegraphics[width=0.6\linewidth]{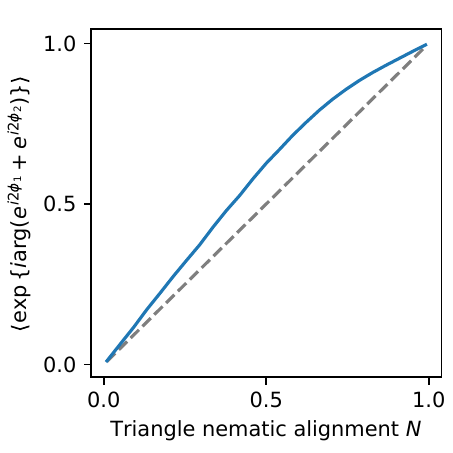}
    \caption{
    Averaging effect in kites.
    Numerical integration of Eq.~\eqref{eq:kite_averaging} for different values of the single-triangle nematic order $N$.
    }
    \label{SI-fig:kite_averaging}
\end{figure}

To carry out the average over $\phi$ in the coarse-graining calculation, it is convenient to use complex numbers to represent traceless $2\times 2$ tensors such as $\mathbf{N}$, with the absolute value indicating the tensor norm and argument indicating tensor orientation.
We assume the distribution $\mathcal{P}(\phi)$ of triangle orientations is a wrapped normal distribution with standard deviation $\sigma$ and mean $\mu$.  
This assumption is justified due to the random walk dynamics of $\phi$. The circular moments are $\overbar{e^{im\phi}}=e^{in\mu-m^2\sigma^2/2}$. In particular, $\overbar{\mathbf{N}}= \overbar{e^{i2\phi}}$. Hence, $\sigma$ is fixed by the nematic order: $\sigma^2 = -\tfrac{1}{2}\log \overbar{N}$.
This allows carrying out the integral Eq.~\eqref{eq:kite_averaging} over the triangle orientations $\phi, \phi'$ in a kite to obtain the mean T1 strain rate. This integral is evaluated numerically (see Fig.~\ref{SI-fig:kite_averaging}). Fitting with a quadratic function results in Eq.~\eqref{eq:kite_averaging}.

The same technique can be used to derive Eq.~\eqref{eq:n_dot_biased}, the change in nematic order due to biased tension feedback. Starting from Eq.~\eqref{eq:phi_dot_biased}, we have
\begin{align}
    \tau \partial_t \overbar{\mathbf{N}} &= \tau \partial_t \overbar{e^{i2\phi}} =  2i\overbar{e^{i2\phi} \partial_t {\phi}} \\
    & \approx -\frac{N_{S}}{2\bar{q}} \overbar{ e^{i2\phi} (e^{i2(\phi - \phi_S)} - e^{-i2(\phi - \phi_S)})} \nonumber
    \\
    &= \frac{N_{S}}{2\bar{q}} e^{i2\phi_S} (1- \overbar{e^{i4(\phi-\phi_S)}})  \nonumber \\
    &= \frac{N_{S}}{2\bar{q}} e^{i2\phi_S} (1- \overbar{N}^4 e^{i4(\mu-\phi_S)}) \nonumber \\
    &\approx \frac{N_{S}}{2\bar{q}} e^{i2\phi_S} (1- \overbar{N}^4)
\end{align}
where in the second step, we have replaced $q$ by its expectation $\bar{q}$, and in the last step, we assumed that if $\overbar{N}^4$ is appreciable, i.e.\ $\overbar{N} \sim 1$, the nematic orientation will be aligned with the source.

\subsection{Maximum entropy parametrization for $\gamma(\bar{q})$}{\label{app:max_ent}}

As noted in the main text, the T1-flux integral Eq.~\eqref{eq:gamma_flux_integral} in principle depends on the details of the probability distribution $\mathcal{P}$, but we want closed evolution equations for the first moments $\bar{q}, \overbar{\mathbf{N}}$ only. To this end, we parameterize $\mathcal{P}$ by the mean using the maximum entropy method. We maximize the entropy $\int_{q,\psi} \mathcal{P} \log\mathcal{P} q dq d\psi$ under the constraints that (1) the marginal of $\psi$ is uniform, (2) the mean of $q$ is $\bar{q}$, (3) $q$ always lies in the ``legal'' region of shape space, $q_0\leq q \leq q_\mathrm{T1}(\psi)$. In line with standard results, the maximum entropy distribution has an exponential form:
\begin{align}
    \mathcal{P}_\mathrm{MaxEnt}(q,\psi) =
    \begin{cases} \frac{e^{\lambda_q q}}{Z(\lambda_q, \psi)}
    & \text{if }  q_0\leq q\leq q_\mathrm{T1}(\psi) \\ 0 & \text{else} \end{cases}
    \label{eq:max_ent_P}
\end{align}
where the normalization factor reads
\begin{align}
    Z(\lambda_q, \psi) = \pi \frac{e^{\lambda_q q_\mathrm{T1}(\psi)}(\lambda_q q_\mathrm{T1}(\psi) -1) - e^{\lambda_q q_0}(\lambda_q q_0 -1)}{\lambda_q^2}
\end{align}
The Lagrange multiplier $\lambda_q$ determines the mean $\bar{q}$. We can solve for $\lambda_q$ as a function of $\bar{q}$, and then insert Eq.~\eqref{eq:max_ent_P} into the T1 flux integral Eq.~\eqref{eq:gamma_flux_integral} to (numerically) obtain $\gamma(\bar{q})$ as shown in Fig.~\ref{fig:4}A. The explicit form for $\gamma(\bar{q})$ is fairly involved. By expanding, we find that it diverges as $\overbar{q}\to\overbar{q_\mathrm{T1}}$ as 
\begin{align}
    \tilde{\tau}\gamma(\bar{q}) = \frac{1}{1- \bar{q}/\overbar{q_\mathrm{T1}}} + \dots
\end{align}
The divergence is due to the fact that the entire probability mass gets ``pushed'' towards the T1 threshold as $\overbar{q}\mapsto\overbar{q_\mathrm{T1}}$, and so the flux across the threshold rapidly increases.

\subsection{Triangle shape change contribution to strain rate}

In the main text, we have exploited the direct connection between cell centroids and the tension triangulation in the Voronoi model to derive Eq.~\eqref{eq:U_dot} for the tissue-scale strain rate. Eq.~\eqref{eq:U_dot} only contains a term taking into account shape change by neighbor exchanges. Tissue can, however, also change shape by cell deformation in the absence of neighbor exchanges, giving rise to an additional term \cite{Blanchard.etal2009}. By Eq.~\eqref{eq:LTC_definition}, the cell shape anisotropy equals $q\mathbf{N}$ \footnote{At this point, it may be useful to compare our definition of the LTC parameters with the results in Ref.~\cite{Merkel.etal2017}, which presents a systematic approach to ``tissue tectonics'' based on the cell centroidal triangulation. Note that the definition of triangle anisotropy in Ref.~\cite{Merkel.etal2017} i.e.\ the eigenvalue of the triangle shear tensor) equals $\frac{1}{4} (\log(1+q)-\log(1-q)) \approx q/2 + O(q^3)$.}.

Because we assume triangles do not rotate in the absence of T1s, the extra contribution to the strain rate is $\partial_t\overbar{\mathbf{U}}_\mathrm{shape} = \overbar{\mathbf{N}} \partial_t \bar{q}$. It is strongly dominated by the T1-driven strain rate $|\partial_t\overbar{\mathbf{U}}_\mathrm{T1}| \gg |\partial_t\overbar{\mathbf{U}}_\mathrm{shape}|$ and we neglect it in the main text -- cells cells would need to become extremely deformed for $\overbar{\mathbf{U}}_\mathrm{shape}$ to become large. Its integrated contribution can be bounded from above by $\log(\overbar{U}_\mathrm{shape}) = \overbar{N}(0)(\bar{q}_* - \bar{q}(0))$. This upper bound shows that $\bar{q}(0)$ has a weak negative impact on total tissue shape change (see Fig.~\ref{fig:3}C).
$\overbar{\mathbf{U}}_\mathrm{shape}$ becomes important when cells can become more highly deformed due to a shifted T1 threshold, e.g.\ in simulations with pinned boundaries (see Fig.~\ref{SI-fig:fixed_boundaries}).

\section{Tension dynamics due to strain-rate feedback}{\label{app:strain_rate_feedback}}

In this and previous work, we have assumed that tissue is in approximate mechanical equilibrium during morphogenesis, since morphogenesis occurs over a much slower time scale than mechanical relaxation.
However, in a network where forces are controlled actively, mechanical relaxation is not sufficient to ensure convergence to mechanical equilibrium -- active feedback loops are necessary. If the network of tensions, i.e.\ the tension triangulation, is not planar, no positions of the cell vertices can satisfy force balance (e.g., consider increasing the active tensions on all edges of a cell by a large amount).
Ref.~\cite{Noll.etal2017} proposes such a feedback loop for tension-dominated epithelial sheets. The active tension $T_{ij}$ on an interface of length $\ell_{ij}$, controlled by the density of motor molecules, is coupled to the strain rate:
\begin{align}
    \frac{\partial_t T_{ij}}{T_{ij}} = \alpha \frac{\partial_t \ell_{ij}}{\ell_{ij}}
    \label{eq:strain_rate_feedback}
\end{align}
where $\alpha$ is a proportionality coefficient. This ensures convergence to mechanical equilibrium without fixing the absolute values of $T_{ij}$ or $\ell_{ij}$.
In Ref.~\cite{Gustafson.etal2022}, optogenetic experiments and mutant analyses were used to show that the effect described by Eq.~\eqref{eq:strain_rate_feedback} is present in the early \emph{Drosophila} embryo. In particular, it was shown that the strain rate induced in the germ band by the invagination of the adjacent ventral furrow is the main factor in setting up the initial anisotropic tension pattern in the germ band prior to convergent extension.

Crucially, Eq.~\eqref{eq:strain_rate_feedback} implies that if the junction lengths $\ell_{ij}$ are subjected to a large-scale shear transformation, i.e.\ $\partial_t {\ell}_{ij} = \ell_{ij} \mathbf{e}_{ij}^\mathrm{T} \cdot \mathbf{S}\cdot \mathbf{e}_{ij}$
where $\mathbf{e}_{ij}$ are the junction orientations and $\mathbf{S}$ is the shear matrix, the tension triangulation is simply sheared by $\alpha \mathbf{R}(\pi/2)\cdot \mathbf{S} \cdot  \mathbf{R}(\pi/2)^\mathrm{T}$. The rotation $\mathbf{R}(\pi/2)$ is due to the fact that edges of tension triangulation and real-space cell edges are orthogonal.
Because Eq.~\eqref{eq:n_dot_biased} for tension feedback coupled to an external nematic source has the form of a shear transformation of tension triangles, as noted in the main text, it describes the effect of strain-rate feedback due to large-scale tissue shear. The feedback strength is given by $N_{S} = 2\alpha S$.
The shear rate and hence $N_{S}$ is, in general, time-dependent, as is the case for the \emph{Drosophila} germband, where the strain rate created by the ventral furrow is transient. This can be easily incorporated into Eq.~\eqref{eq:n_dot_biased}.

\section{Effect of isogonal strain on distribution of triangle orientation}{\label{app:isogonal_fokker_planck}}

Here, we derive the orientation bias in the presence of isogonal strain from a Fokker--Planck equation.

\subsection{Transformation of interface lengths under isogonal strain}

\begin{figure}[t]
    \centering
    \includegraphics[width=0.6\linewidth]{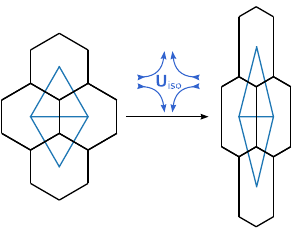}
    \caption{
    Transformation of interface length under isogonal strain.
    Under isogonal strain, interfaces (black) only change length, not orientation, while the vectors connecting cell centroids (blue) change both length and orientation. As a consequence, interface length changes are amplified by a factor of two.  
    }
    \label{SI-fig:isogonal_length}
\end{figure}

We begin by computing how an isogonal transformation changes cell-cell interface lengths. The degree of isogonal strain is measured by the isogonal strain tensor $\mathbf{U}_\mathrm{iso}$ which maps the tension triangulation to the triangulation of cell centroids (in the Voronoi reference state, the centroidal triangulation equals the tension triangulation). $\phi_\mathrm{iso}$ is the orientation of the principal axis of $\mathbf{U}_\mathrm{iso}$.
Under an isogonal transformation, the length $\ell$ of an interface with orientation $\mathbf{e}$ is changed by
\begin{align}
    \Delta\ell_\mathrm{iso}/\ell_0 \approx 2\, \mathbf{e}^\mathrm{T}\cdot \mathbf{U}_\mathrm{iso} \cdot \mathbf{e}
    \label{SI-eq:isogonal_length_change}
\end{align}
where $\ell_0$ is the average Voronoi interface length. Note the additional factor $\times 2$: interface lengths transform differently than under standard shear transformations. Under isogonal strain, cell-cell interfaces (black) only change length, not orientation, while the vectors connecting cell centroids (blue) change both length and orientation.
Intuitively, lengths must change more to obtain the prescribed deformation of cell centroids. This is illustrated in Fig.~\ref{SI-fig:isogonal_length}. For the case of an isosceles tension triangle, a direct geometric calculation confirms Eq.~\eqref{SI-eq:isogonal_length_change}.

\subsection{Shift of T1 threshold by isogonal strain}

The key effect of the isogonal strain lies in shifting when interfaces collapse, in an orientation-dependent manner.
To compute $q_\mathrm{T1}(\phi)$, we make use of the methods introduced in Refs.~\cite{Brauns.etal2024,Claussen.etal2023}. 

Since an isogonal transformation changes interface lengths according to Eq.~\eqref{SI-eq:isogonal_length_change}, T1-transitions now occur at $\ell +\Delta \ell_\mathrm{iso}= 0$. Ref.~\cite{Brauns.etal2024} calculated the shifted T1 threshold in the $(q,\psi)$--plane as a function of $\Delta \ell_\mathrm{iso}/\ell_0$ (see Fig.~\ref{fig:6}B). We now average these results over $\psi$. Since the edge of maximum tension in a triangle is (approximately) oriented along $\phi$, $\Delta \ell_\mathrm{iso} /\ell_0= 2 U_\mathrm{iso} \cos( 2 (\phi-\phi_\mathrm{iso}))$. We obtain
\begin{align}
    q_\mathrm{T1}(\phi) \approx q_\mathrm{T1}^{(0)} + q_\mathrm{T1}^{(1)} U_\mathrm{iso} \cos 2 (\phi-\phi_\mathrm{iso}) +\mathcal{O}(U_\mathrm{iso}^2)
    \label{eq:q_T1_isogonal}
\end{align}
where the numerical constants are found to be $q_\mathrm{T1}^{(0)}\approx 0.6, \; q_\mathrm{T1}^{(1)}\approx 0.6$.
We find good agreement between this formula and the numerical simulations with anisotropic reference shape. From Eq.~\eqref{eq:q_T1_isogonal} and Eq.~\eqref{eq:tau_T1}, we can compute the steady-state rate of T1s $\tau_\mathrm{T1}^{-1}=(\log q_\mathrm{T1}(\phi)/q_0)^{-1}$ (see Appendix~\ref{app:steady_state}).

\subsection{Fokker--Planck equation for triangle orientation}

The dynamics of $\phi$ in the stochastic model for LTC dynamics presented above is defined by random re--orientations $\phi \mapsto \phi +\delta\phi, \; \delta\phi \sim \mathcal{N}(0,\sigma_\phi^2)$ that occur each time the triangle hits the T1 threshold. Hence, $\phi$ executes a random walk with diffusion constant $\sigma_\phi^2/\tau_\mathrm{T1}$, where $\tau_\mathrm{T1}=\tilde{\tau}\log q_\mathrm{T1}(\phi)/q_0$ is the time between T1s.
This observation allows us to simplify the Ito Fokker--Planck equation for the
marginal distribution $\mathcal{P}(\phi)$:
\begin{align}
    \label{eq:fokker_planck_isogonal}
    \partial_t \mathcal{P}(\phi) = \frac{1}{2}\partial_{\phi}^2 (\sigma_\phi^2\tau_\mathrm{T1}^{-1} \mathcal{P})
\end{align}
In the presence of isogonal strain, the diffusion constant becomes orientation-dependent. Note that the usual argument that a gradient in the diffusion constant cannot create a density gradient does not apply here; the dynamics of $\phi$ does not satisfy detailed balance because of cycles in the full $(q, \psi, \phi)$ space. Numerical simulation of the mean-field model confirms that the Fokker--Planck equation \eqref{eq:fokker_planck_isogonal} is correct.

First, as argued above the T1-threshold anisotropy becomes orientation dependent, $q_\mathrm{T1}= q_\mathrm{T1}(\phi)$, and hence $\tau_\mathrm{T1} = \tau_\mathrm{T1}(\phi)$. As a consequence, the amount of orientation change $\sigma_\phi$  orientation--dependent as well: more anisotropic triangles are reoriented less strongly by edge flips.

As noted, also the second factor in the diffusion coefficient, $\sigma_\phi(\phi)^2$ becomes orientation dependent. This is because the degree of reorientation of a triangle by a T1 depends on its anisotropy at the moment of the T1. As we explain in the main text and in Appendix~\ref{app:coarse_graining}, $\sigma_\phi$ is approximately given by the angle between the longest and second-longest edge in a triangle at the moment of the T1. Hence, the more anisotropic the triangle becomes, the less its orientation is affected by T1s. Following the approach described above and averaging $\sigma_\phi$ over $\psi$ along the shifted T1 curves of Fig.~\ref{fig:6}B, we obtain to second order in $U_\mathrm{iso}$
\begin{equation}
    \sigma_\phi(\phi) \approx \sigma_\phi^{(0)} + \sigma_\phi^{(1)} U_\mathrm{iso} \cos 2 (\phi_\triangle-\phi_\mathrm{iso})
\end{equation}
with numerical values of $\sigma_\phi^{(0)}\approx 0.2\pi, \; \sigma_\phi^{(1)}\approx 0.18\pi$. 

In the steady state, we then have
\begin{align}
    \partial_{\phi}^2 ( \sigma_\phi^2\tau_\mathrm{T1}^{-1} \mathcal{P}) \stackrel{!}{=} 0 \quad \Rightarrow \quad \mathcal{P} \propto \sigma_\phi(\phi)^{-2} \tau_\mathrm{T1}(\phi) 
    \label{eq:isogonal_angle_dist}
\end{align}
Hence, the shift in the T1-threshold creates a bias in triangle orientation. Calculating the expectation value of $N$ in the steady state gives:
\begin{equation}
    \overbar{N}_* \approx 1.5 U_\mathrm{iso}.
\end{equation}

The T1-rate as a function of $\phi$ also acquires a bias: $\gamma = \mathcal{P}(\phi) / \tau_\mathrm{T1}(\phi) \propto \sigma_\phi(\phi)^2$. From this and Eq.~\eqref{eq:U_dot}, we can calculate the predicted strain rate shown in Fig.~\ref{fig:6}G.  Hence, we expect net convergent extension, along the axis of isogonal strain. In the full numerical simulations, we implemented isogonal strain by using an anisotropic reference shape, as discussed in the simulation methods.

\begin{figure*}
    \centering
    \includegraphics[width=0.9\textwidth]{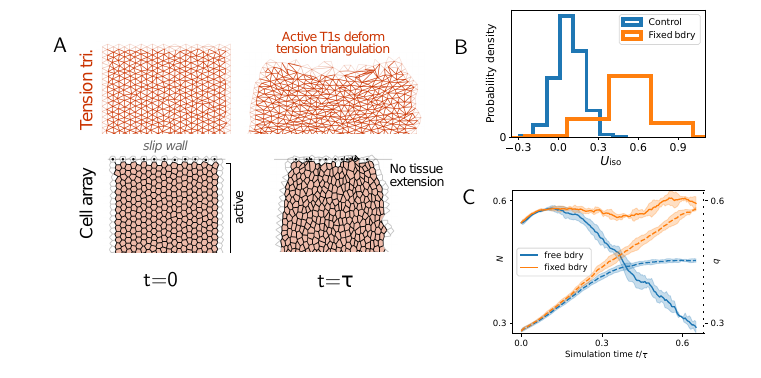}
    \caption{
    Simulations with fixed boundaries.
    \textbf{A}~Simulations with fixed boundaries, implemented via additional energetic penalty on cells marked with black dots. Adapted from Ref.~\cite{Claussen.etal2023}. Only one-half of a top-bottom symmetric simulation domain is shown
    \textbf{B}~Dynamics of $q, N$ compared with free-boundary control simulations show the impact of shifted T1 threshold (increased $q, N$).
    \textbf{C}~Distribution of quartet-level isogonal strain $\mathbf{U}_\mathrm{iso}$ anisotropy for fixed and free boundary simulations. $\mathbf{U}_\mathrm{iso}$ was computed as the map from Voronoi to physical cell centroids.}
    \label{SI-fig:fixed_boundaries}
\end{figure*}

\section{Simulation methods}{\label{app:simulation_methods}}

To perform out and analyze the cell-resolved simulations presented in this work, we used the numerical model and code previously described in Ref.~\cite{Claussen.etal2023}. The source code for these simulations is available on GitHub \cite{Claussen2023}. 
In brief, this code describes a tissue by a set of planar tension triangulation vertices $\mathbf{t}_i$ (once for each cell), defining the interfacial tensions as $T_{ij} = |\mathbf{t}_i-\mathbf{t}_j|$, and tri-cellular vertices $\mathbf{r}_{ijk}$. Time evolution is based on the following algorithm at each timestep $t\mapsto t+dt$:
\begin{table}[H]
\begin{enumerate}
  \item Compute (infinitesimal) tension changes $dT_{ij} = \tau^{-1} f(T_{ij}, T_{ij,p}) dt$ according to Eq.~\eqref{eq:tension-dyn}.
  \item Move the tension vertices so that $|\mathbf{t}_i-\mathbf{t}_j| \approx T_{ij} + dT_{ij}$ using least-squares to minimize the deviation $\sum_{(ij)} \big[|\mathbf{t}_i-\mathbf{t}_j| - (T_{ij} + dT_{ij})\big]^2$.
  \item Update the tensions using the new tension triangulation $T_{ij} = |\mathbf{t}_i-\mathbf{t}_j|$.
  \item Find cell vertex positions $\mathbf{r}_{ijk}$ in physical space by minimizing the elastic energy \eqref{eq:total-energy}.
  \item Carry out topological modifications for any collapsed interface $\ell_{ij} \leq 0$. 
\label{alg:tissue-sim}
\end{enumerate}
\caption{Numerical time evolution in full cell-resolved simulations. See Ref.~\cite{Claussen.etal2023} for details.}
\end{table}

For the subdominant cell shape energy appearing in \eqref{eq:total-energy} we use
\begin{align}
   E_\mathrm{shape}(\mathbf{S}) &= \lambda [\mathrm{Tr}(\mathbf{S}-\mathbf{S}_0)]^2 + \mu \mathrm{Tr}[(\mathbf{S}-\mathbf{S}_0)^2] \\
   \mathbf{S}_{i} &= \sum_{k \in \mathcal{N}_i} \frac{ \mathbf{r}_{ik} \otimes \mathbf{r}_{ik}}{\ell_{ik}},
   \label{eq:cell_shape_energy}
\end{align}
where $\mathcal{N}_i$ is the set neighbors of cell $i$, and $\mu,\lambda$ are elastic moduli, and $\mathbf{S}_0$ is the reference shape. Dynamics arises via the adiabatic changes of the tension triangulation, as described in the main text.

To implement the tension-driven Voronoi model, we modified the numerical model of Ref.~\cite{Claussen.etal2023} by setting the cell vertex positions at each time step to the Voronoi centroids of the tension triangles, instead of obtaining them by minimization of the elastic energy. The relevant code is part of the GitHub repository \cite{Claussen2023}. For clarity, we denote simulations where vertex positions were obtained by elastic energy minimization rather than the Voronoi prescription as ``elastic energy'' simulations.

The majority of elastic energy simulations studied here were carried out and previously analyzed in Ref.~\cite{Claussen.etal2023}, in particular in the ``phase diagram'' of convergent extension as a function of the initial tension geometry (see Fig.~2 of Ref.~\cite{Claussen.etal2023}), as well as the supplementary information (Fig.~S7 of Ref.~\cite{Claussen.etal2023}). We always used the default parameter values described in Table~1 of Ref.~\cite{Claussen.etal2023}, which were obtained by a fit to data on \emph{Drosophila} germ band extension.

All comparisons of the theory with numerics in the present work were done using the full elastic energy simulations.

\subsection{Initial conditions for tissue scale simulations}

For the creation of initial conditions --i.e.\ the initial tension triangulation--  for the simulations, we used two different methods. For the phase diagram simulations previously described in Ref.~\cite{Claussen.etal2023}, we used a hard-disk-packing based method, as explained in the main text. In brief, we sampled tension vertex positions $\mathbf{t}_i$ from a hard disk packing process \cite{Bernard.etal2009} with a packing fraction $\rho\in [0, 1]$ controlling the degree of order. $\rho=0$ corresponds to Poisson deposition, and $\rho=1$ to crystalline hexagonal packing. The tension triangulation is obtained as the Delaunay triangulation of the sampled point set. For all other simulations (see Figs.~ \ref{fig:1}, \ref{fig:tension-source}, \ref{fig:6}, \ref{SI-fig:assumptions_check}, \ref{SI-fig:fixed_boundaries}, and 
\ref{SI-fig:Voronoi_vs_elastic}), we used a perturbed triangular lattice as an initial condition for the tension triangulation. Next, the triangulation vertices were subjected to a shear transformation $\mathbf{t}_i\mapsto \mathrm{Diag}(1+s/2, 1-s/2)\mathbf{t}_i$ to tune the initial tension anisotropy. 

\subsection{Implementation of anisotropic cell reference shapes}

To implement persistent anisotropic cell shapes, we used elastic energy simulations with an anisotropic reference cell shape tensor $\mathbf{S}_0$, as previously described \cite{Claussen.etal2023}. $\mathbf{S}_0$ directly controls the anisotropic isogonal mode: we find that the relative anisotropy of $\mathbf{U}_\mathrm{iso}$ is approximately equal to that of $\mathbf{S}_0$ after energy minimization.

\subsection{Comparison of tension-driven Voronoi and full elastic energy simulations}

\begin{figure*}
    \centering
    \includegraphics[width=\textwidth]{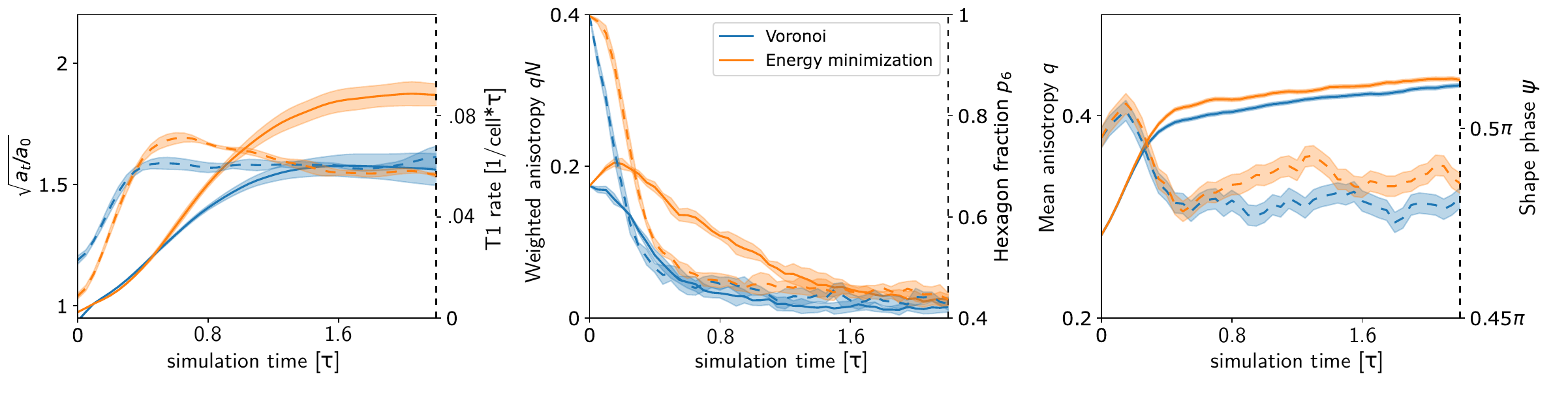}
    \caption{
    Comparison of tension-driven Voronoi and elastic energy minimization models.
    In the TDV model, the cell shapes are determined via Voronoi--Delaunay duality from the tension triangulation (blue) while in the ``full'' model where cell shapes are obtained by minimizing the cell shape energy under the angle constraints imposed by the tension triangulation (orange). Simulations were initialized with an ``ordered'' tension triangulation (generated from hard disks at high packing fraction $\rho = 0.9$) and initial tension anisotropy magnitude $s=0.2$.
    Note that we plot here the weighted anisotropy $\overbar{qN}$ instead of $\overbar{N}$ as in the remainder of this manuscript to facilitate comparison with the simulation plots of Ref.~\cite{Claussen.etal2023}.
    }
    \label{SI-fig:Voronoi_vs_elastic}
\end{figure*}

Here, we compare the TDV model to ``full'' numerical simulations in which cell shape is obtained by direct minimization of the elastic energy Eq.~\eqref{eq:total-energy}. The dynamics of the tensions Eq.~\eqref{eq:tension-dyn} is identical between the two models. As shown by Fig.~\ref{SI-fig:Voronoi_vs_elastic}, the Voronoi simulations behave quantitatively quite similar to the full simulations. The observer difference in total elongation is due to the fact that cell shapes in the Voronoi model start already elongated along the $x$-axis, due to Voronoi-Delaunay duality and the initial anisotropy in the tension triangulation.

Beyond the numerics, there are also theoretical arguments for the agreement between the full and Voronoi models. First, regarding the instantaneous tissue geometry, both the Voronoi and the full model respect tension force balance, and hence the angles at tri-cellular vertices agree between the two \cite{Claussen.etal2023}. The only difference lies in the residual isogonal mode (one degree of freedom per cell). The Voronoi construction (associate each point in the plane to the nearest triangulation vertex) tends to lead to isotropic cells of relatively similar size, leading to similar results than minimization of a cell shape elastic energy. For very heterogeneous triangle sizes, and very acute triangles, this may no longer be true; however, these do not occur in our simulations.

Second, the dynamics of the model is driven by the deformation of the tension triangulation. The only way the real-space cell shapes (and hence the difference between full and Voronoi simulations) feed into the tension dynamics is by determining when T1s occur. And at least in a symmetric cell quartet \cite{Claussen.etal2023}, full elastic energy minimization produces the same result as the Delaunay criterion that governs T1s in the Voronoi simulations.

As explained in the main text, however, the Voronoi simulations are incapable of incorporating external forces, e.g.\ due to fixed boundary conditions.

\bibliography{GBE.bib}

\end{document}